\documentclass{article}
\usepackage{arxiv}

\usepackage[normalem]{ulem} 

\usepackage[utf8]{inputenc} 
\usepackage[T1]{fontenc}    
\usepackage{hyperref}       
\usepackage{url}            
\usepackage{booktabs}       
\usepackage{amsfonts}       
\usepackage{microtype}      
\usepackage{color}
\usepackage{lipsum}
\usepackage{amsthm}
\usepackage{indentfirst}
\usepackage{graphicx}
\usepackage{multicol}
\usepackage{epstopdf}
\usepackage{fourier}
\usepackage{bm}
\usepackage{bbm} 
\usepackage{mathtools}
\usepackage{latexsym,enumerate}
\usepackage{ulem}
\usepackage{subfig}

\newcommand{\comment}[1]{}

\newcommand{\BEA}{\begin{eqnarray}}
\newcommand{\EEA}{\end{eqnarray}}

\title{Machine Learning-Based Statistical Closure Models for Turbulent Dynamical Systems}

\author{
 Di Qi \\
  Department of Mathematics \\
  Purdue University, West Lafayette, IN 47907, USA \\
  \texttt{qidi@purdue.edu} \\
  \And
  John Harlim \\
  Department of Mathematics, Department of Meteorology and Atmospheric Science, \\ Institute for Computational and Data Sciences \\
  The Pennsylvania State University, University Park, PA 16802, USA\\
  \texttt{jharlim@psu.edu} \\
}

\begin{document}

\maketitle

\begin{abstract}
We propose a Machine Learning (ML) non-Markovian closure modeling framework for accurate predictions of statistical responses of turbulent dynamical systems subjected to external forcings. One of the difficulties in this statistical closure problem is the lack of training data, which is a configuration that is not desirable in supervised learning with neural network models. In this study with the 40-dimensional Lorenz-96 model, the shortage of data is due to the stationarity of the statistics beyond the decorrelation time. Thus, the only informative content in the training data is from the short-time transient statistics. We adopt a unified closure framework on various truncation regimes, including and excluding the detailed dynamical equations for the variances. The closure framework employs a Long-Short-Term-Memory architecture to represent the higher-order unresolved statistical feedbacks with a choice of ansatz that accounts for the intrinsic instability yet produces stable long-time predictions. We found that this unified agnostic ML approach performs well under various truncation scenarios. Numerically, it is shown that the ML closure model can accurately predict the long-time statistical responses subjected to various time-dependent external forces that have larger maximum forcing amplitudes and are not in the training dataset.  
\end{abstract}


\section{Introduction}

Closure problem in nonlinear dynamical systems is one of the most challenging tasks in computational statistics, see e.g.,
\cite{lesieur1987turbulence,kwasniok2012data,lu2016comparison,SapsisMajda2013,qi2016low,maulik2019subgrid}. In the context of turbulent fluid flows, closure problem has been studied for over a century dated back to Boussinesq's eddy viscosity hypothesis \cite{SCHMITT2007617}, where the goal is to describe the Reynold stress term (which is effectively a second-order statistic) as a function of the mean flow. In a nutshell, the underlying closure problem is to find a closed system that can describe the evolution of observable (such as low-order statistics), and by ``closure'', the goal is to specify a map that allows one to untangle the dependence on unresolved variables (such as higher-order statistics). In the context of low-order statistical closure problem, which is the primary interest in this work, predicting the time evolution of mean statistics is useful for point estimation, while predicting the time evolution of the covariance statistics has a wide range of applications, including uncertainty quantification \cite{leith1975climate,majda2018strategies,majda2019linear} and data assimilation \cite{bh:14,h:15,zh:15}.

As machine learning becomes popular, finding such a ``closure'' map can be formulated as a supervised learning task. With machine learning algorithms, one approximates the closure system by solving a regression problem on an appropriate hypothesis space, replacing the traditional approach of finding an analytical expression that can be very difficult in general. In the context of turbulent fluid flows, numerous neural network-based machine learning closure systems have been proposed (see e.g., \cite{gamahara2017searching,singh2017machine,maulik2019subgrid}). While the success of the estimation depends crucially on the choice of neural network architectures, a natural hypothesis space for modeling time series is the Recurrent Neural Networks (RNNs) architecture. In the closure modeling applications, the Long-Short-Term-Memory (LSTM) \cite{Hochreiter_1997}, a special class of RNNs, has been shown to produce state-of-art accuracies in the prediction of high-dimensional time series \cite{MWE:2018,vlachas2018data,maulik2020time,HJLY:19}. 

Building on these empirical successes, we consider the LSTM-based neural network architecture for statistical closure modeling of turbulent dynamical systems. In this paper, we will examine the effectiveness of machine learning in uncovering the non-Markovian statistical model. In previous works \cite{jh:20,HJLY:19}, a closure model for predicting the trajectory of the observed state variables is constructed using a long time series of the corresponding observable. Despite the similarity to the closure modeling framework formulated in \cite{jh:20,HJLY:19}, the proposed statistical closure problem in this article is more challenging. In the present work, a unified model framework is proposed aiming to directly predict the leading-order statistical moments subjected to general external perturbations, with limited training data. Particularly, we will consider short-time transient statistical sequences for training. This consideration is motivated by practical issues (e.g., stiff numerical solver and large storage) in obtaining longer time series when the full-order model is multiscale and high-dimensional. Even in moderately low-dimensional problems, as we shall see in this paper when the perturbed dynamics correlation statistics are decaying, we only have short time series of transient statistics that are informative for training. While the lack of training data makes the closure problems in this paper a stringent test for the machine learning algorithm, we are not only concerned to predict the evolution of the low-order statistics of the underlying unperturbed system. Our ultimate goal is to capture statistical responses subjected to unseen external forces, extending previous works \cite{jh:20,HJLY:19} which only examined the accuracy of the unperturbed dynamical system on new initial conditions. 

To achieve this goal, we assume that one can numerically simulate the full-order model in a short time window (as in many reduced-order modeling configurations, e.g., \cite{givon2004extracting,eweinan2003CMS,chk:02,gouasmi2017priori,majda2018strategies}) to generate a training dataset under pre-selected simple constant forcing functions and initial conditions. We will simulate this training dataset by a Monte-Carlo simulation. While this task can be expensive depending on: the choice of integration scheme for solving the underlying full-order model, the length of the time integration to reach correlation time scales, and the sample size needed to achieve a robust statistical estimation, it only needs to be performed once for pre-selected constant external forcings. Subsequently, we validate the closure model by examining how well it can extrapolate beyond the training data to predict the statistical responses subjected to various new time-dependent forcing functions and initial conditions.

Numerically, we examine the machine learning closure on a simple test model, the Lorenz '96 (L-96) system, that was first introduced by Lorenz \cite{lorenz:96} as an idealization of atmospheric waves in midlatitude. While the model is simple, it carries some properties of realistic turbulent complex systems \cite{majda2016introduction,majda2016improving} such as the energy preserving advection-like nonlinear term, and a wide spectrum of unstable modes through the nonlinear coupling between states. Beyond the simplicity, which allows us to carry the numerical verification with moderate computational costs, our choice to investigate this case is largely motivated by the fact that closure models for a coupled system of the mean and covariance statistics have been well-developed and improved in \cite{SapsisMajda2013,majda2016improving,majda2018strategies}. These parametric closure models, developed based on clever physical intuition, have demonstrated accurate statistical predictions. In such a configuration, we found that the machine learning-based model can produce accurate statistical responses (comparable to the parametric model) on moderate to large forcing amplitudes. 
Despite the effective prediction with parametric closure models in \cite{majda2016improving,majda2018strategies}, the cost in calibrating the statistical modes throughout the entire spectrum can become very expensive as the dimensionality of the problem increases \cite{majda2019linear,qi2016low}. In addition to this practical problem, a more fundamental issue with parametric modeling is that the design of accurate closure models crucially depends on knowing the physics well enough, such as self-similarity or some structure of the underlying dynamics. As an example that illustrates this issue, we will compare parametric and ML closure models for only the mean statistics (no dynamical models for the variance are involved) in the simple L-96 example. In this scenario, we find that the agnostic ML framework can produce more accurate predictions, beating the parametric-based approach. This simple test suggests that the agnostic approach is easily portable for any truncation scenario. On the other hand, while the parametric modeling assumption \cite{majda2016improving,majda2018strategies} works well on the coupled system of the mean of covariance statistics, different parametric assumptions need to be considered for accurate closure of only the mean statistics.

The remainder of this paper is organized as follows. In Section~\ref{sec2}, we discuss the general statistical closure modeling framework of turbulent dynamical systems using L-96 as a prototypical example and provide a hierarchy of low-order closure models. In Section~\ref{sec3}, we provide details on the machine learning algorithm used to estimate the non-Markovian dynamical components. In Section~\ref{sec4}, we present numerical results on the hierarchy of closure models introduced in Section~\ref{sec2}. In Section~\ref{sec5}, we close the paper with a summary.

\section{Statistical closure of complex nonlinear systems}\label{sec2}

The general formulation of the turbulent dynamical
systems \cite{majda2016introduction,majda2018strategies} can be described by the canonical equations for the state
variable $\mathbf{u}\in\mathbb{R}^{N}$ as,
\begin{equation}
\frac{d\mathbf{u}}{dt}=\left(\mathcal{L+D}\right)\mathbf{u}+B\left(\mathbf{u},\mathbf{u}\right)+\mathbf{F}\left(t\right).\label{eq:abs_formu}
\end{equation}
On the right hand side of the above equation (\ref{eq:abs_formu}),
the first two components, $\left(\mathcal{L+D}\right)\mathbf{u}$,
represent linear dispersion and dissipation effects, where $\mathcal{L}^{*}=-\mathcal{L}$
is an energy-conserving skew-symmetric operator; and $\mathcal{D}<0$
is a negative definite operator. The nonlinear effect in the dynamical
system is introduced through a quadratic form, $B\left(\mathbf{u},\mathbf{u}\right)$,
that satisfies the conservation law, $\mathbf{u}\cdot B\left(\mathbf{u},\mathbf{u}\right)=0$, 
and the Liouville property, $\mathrm{div}_{\mathrm{u}}B\left(\mathbf{u},\mathbf{u}\right)=0$ \cite{majda2016introduction}.

Following \cite{sapsis2013statistically,majda2018strategies}, the dynamics of the statistical moments are constructed by representing the state space $\mathbf{u}$ as,
\begin{equation}
\mathbf{u}\left(t\right)=\bar{\mathbf{u}}\left(t\right)+\sum_{i=1}^{N}Z_{i}\left(t\right)\mathbf{e}_{i},\quad \quad R_{ij} =\left\langle Z_{i}Z_{j}^{*}\right\rangle ,\label{eq:spec_expansion}
\end{equation}
where $\bar{\mathbf{u}}\left(t\right)=\left\langle \mathbf{u}\left(t\right)\right\rangle$
represents the mean statistics, and the coefficients $\left\{ Z_{i}\left(t\right)\right\} $
are fluctuation terms along the coordinates $\mathbf{e}_{i}$.  In the above description, the notation $\langle \cdot \rangle$ is to denote the canonical statistical ensemble average that approximates the integral over the phase space at the limit of large ensemble size, following the standard notion in statistical mechanics \cite{zwanzig2001nonequilibrium}. Inserting the representation in  \eqref{eq:spec_expansion} to \eqref{eq:abs_formu}, one obtains a system of dynamical moments equations, where the first two moments satisfy,
\begin{subequations}
\begin{align}
\frac{d\bar{\mathbf{u}}}{dt} &=\left(\mathcal{L}+\mathcal{D}\right)\bar{\mathbf{u}}+B\left(\bar{\mathbf{u}},\bar{\mathbf{u}}\right)+\sum_{i,j}R_{ij}B\left(\mathbf{e}_{i},\mathbf{e}_{j}\right)+\mathbf{F}, \label{eq:mean_dyn} \\
\frac{dR}{dt} &= L_{v}\left(\bar{\mathbf{u}}\right)R+RL_{v}^{*}\left(\bar{\mathbf{u}}\right)+\theta, \label{eq:cov_dyn}
\end{align}
\end{subequations} 
with
\begin{subequations}
\begin{align}
\left( L_{v}\right) _{ij}&=\left[\left(\mathcal{L}+\mathcal{D}\right)\mathbf{e}_{j}+B\left(\bar{\mathbf{u}},\mathbf{e}_{j}\right)+B\left(\mathbf{e}_{j},\bar{\mathbf{u}}\right)\right]\cdot\mathbf{e}_{i},\label{eq:lin_operator}\\
\left( \theta\right) _{ij}&=\sum_{m,n}\left\langle Z_{m}Z_{n}Z_{j}\right\rangle B\left(\mathbf{e}_{m},\mathbf{e}_{n}\right)\cdot\mathbf{e}_{i}+\left\langle Z_{m}Z_{n}Z_{i}\right\rangle B\left(\mathbf{e}_{m},\mathbf{e}_{n}\right)\cdot\mathbf{e}_{j}.\label{eq:nonlinear_flux}
\end{align}
\end{subequations} 
Here, the energy flux $\theta$ expresses nonlinear energy exchanges between different fluctuation modes due to the nonlinearity of the dynamics modeled through third-order moments. In general, such a representation gives rise to a non-closed system (possibly infinite-dimensional ODEs) as each moment equation is coupled to higher-order moments. 

Despite the fact that the exact equations for the statistical mean
(\ref{eq:mean_dyn}) and the covariance fluctuations (\ref{eq:cov_dyn})
are not a closed system, the total energy in the mean plus the total variance defined as $E = \frac{1}{2}\bar{\mathbf{u}}\cdot\bar{\mathbf{u}}+\frac{1}{2}\mathrm{tr}(R)$
satisfies the following scalar dynamical equation \cite{majda2018strategies},
\begin{equation}
\frac{dE}{dt}=\bar{\mathbf{u}}\cdot \mathcal{D}\bar{\mathbf{u}}+\mathrm{tr}\left(\mathcal{D}R\right)+\bar{\mathbf{u}}\cdot\mathbf{F},\label{eq:energy_conservation}
\end{equation}
where $\bar{\mathbf{u}}$ and $R$ are the exact solutions from the statistical equations. While the mean and covariance dynamics in \eqref{eq:mean_dyn}-\eqref{eq:cov_dyn} are not explicitly written in terms of $E$, we found that by allowing the unresolved components in \eqref{eq:cov_dyn} to depend on $E$, one can achieve an effective non-Markovian closure model, especially for reduced-order model building upon the coupled system \eqref{eq:mean_dyn},\eqref{eq:cov_dyn}, \eqref{eq:energy_conservation} in leading modes \cite{majda2016improving}.
 
For the convenience of notation in the following discussion, we consider a discrete dynamical system induced by numerical integration of the coupled system  \eqref{eq:mean_dyn},\eqref{eq:cov_dyn}, \eqref{eq:energy_conservation}, and a non-Markovian equation for the energy flux $\theta$, 
\begin{equation}
\begin{aligned}\label{discretenonmarkov}
\bar{\mathbf{u}}_{i+1} &= \mathcal{F}_1(\bar{\mathbf{u}}_{i},R_i,\mathbf{F}_{i+1}),\\
R_{i+1} &= \mathcal{F}_2(\bar{\mathbf{u}}_{i},R_i,\theta_i) \\
E_{i+1} &= \mathcal{F}_3(\bar{\mathbf{u}}_{i},R_i,\mathbf{F}_{i+1}) \\
\theta_{i+1} &=\mathcal{G} (\bar{\mathbf{u}}_{i},\ldots,\bar{\mathbf{u}}_{i-m+1};R_i,\ldots, R_{i-m+1}; E_i\ldots, E_{i-m+1}; \theta_{i},\ldots, \theta_{i-m+1}).
\end{aligned}
\end{equation}
Here, we have defined $\bar{\mathbf{u}}_{i}:=\bar{\mathbf{u}}(t_{i})$, $R_i:=R(t_i)$, $E_i: =E(t_i)$, $\theta_i: =\theta(t_i)$, and $\{\mathcal{F}_j\}$ to denote the corresponding operators associated to the numerical integration of \eqref{eq:mean_dyn},\eqref{eq:cov_dyn}, \eqref{eq:energy_conservation} for a suitable time step $\Delta t:=t_{i+1}-t_i$. The operator $\mathcal{G}$ denotes a hidden non-Markovian model that maps the delay coordinates of variables $\{\bar{\mathbf{u}},R,E,\theta\}$ to the energy flux $\theta$ at the next time step. We should point out that in the absence of external forces, the non-Markovian system in \eqref{discretenonmarkov} is an exact representation (no approximation) of the corresponding temporal discretization of the full dynamical system in \eqref{eq:abs_formu} in terms of $\{\bar{\mathbf{u}},R,E,\theta\}$ that satisfies mild conditions of the delay embedding theorem \cite{TAKENS2010345,MR1137425}. To see this, one can employ the discrete Mori-Zwanzig formulation \cite{HJLY:19} to the full system \eqref{eq:abs_formu} with a projection operator defined as the conditional expectation of the delay embedding coordinates of these observables, $\mathbb{E}[\mathbf{X}_{i+1}|\mathbf{x}_{i},\ldots, \mathbf{x}_{i-m+1}]$ for some $m>1$, where $\mathbf{X}_i$ denotes the random variable associated with the dynamical process $\mathbf{x}_i:=(\mathbf{u}_{i},R_i,E_i,\theta_{i})$ (see Section 3 of \cite{gilani2021kernel} for such a derivation). 

While one can, in principle, deduce the hidden dynamics $\mathcal{G}$, such a mathematical derivation is far from trivial even if the structure of the full dynamics in \eqref{eq:abs_formu} is known. Following the idea in \cite{jh:20,HJLY:19}, we will use machine learning to approximate the hidden map $\mathcal{G}$ in an efficient way. Theoretically, under the assumption that $\mathcal{F}_j, \mathcal{G}$ are uniformly Lipschitz, one can guarantee accurate solutions (in a strong sense) up to a finite time with an error bound that depends linearly on the total error of learning $\mathcal{G}$ (see Theorem~3 in \cite{HJLY:19}). Numerically, we will consider a specific type of Recurrent Neural Networks (RNN) known as the Long-Short-Term-Memory model for the estimation of $\mathcal{G}$, motivated by the robust numerical results on other closure problems reported in \cite{HJLY:19}. In fact, using the approximation theory of a two-layer neural network, the work in \cite{levine2021framework} shows that there exists an RNN closure model that gives the desired consistency up to a finite time.

To illustrate the approach, we focus on the Lorenz'96 (L-96) model \cite{lorenz:96} that fits into the general structure of \eqref{eq:abs_formu}. The L-96 model is a 40-dimensional ODE system with state
variables $\mathbf{u}=\left(u_{0},u_{1},...,u_{J-1}\right)^{\top}$
\begin{equation}
\frac{du_{j}}{dt}=\left(u_{j+1}-u_{j-2}\right)u_{j-1}-d\left(t\right)u_{j}+F_{j}\left(t\right),\;j=0,\cdots,J-1=39,\label{eq:L96}
\end{equation}
with a periodic boundary condition, $u_{J}=u_{0}$, mimicking geophysical
waves in the mid-latitude atmosphere. While the model is rather simple, it carries representative properties of realistic complex systems with the energy preserving advection-like nonlinear terms, and the exchanges between the damping and forcing terms.

To compare with the abstract form (\ref{eq:abs_formu}), we
can write the linear and quadratic operators for the L-96 system as
\[
\mathcal{L}= 0, \quad\quad\mathcal{D}\left(t\right)=\mathrm{diag}\left(-d_{0}\left(t\right),\cdots,-d_{J-1}\left(t\right)\right), \quad\quad B\left(\mathbf{u},\mathbf{v}\right)=\left\{ u_{i-1}^{*}\left(v_{i+1}-v_{i-2}\right)\right\} _{i=0}^{J-1}
\]
and define the state variables in \eqref{eq:spec_expansion} with $\mathbf{e}_k := \lbrace e^{2\pi i k\frac{j}{J}}\rbrace_{j}$ for $j=0,\ldots, J-1$.

\begin{figure}
\begin{center}
\includegraphics[scale=0.3]{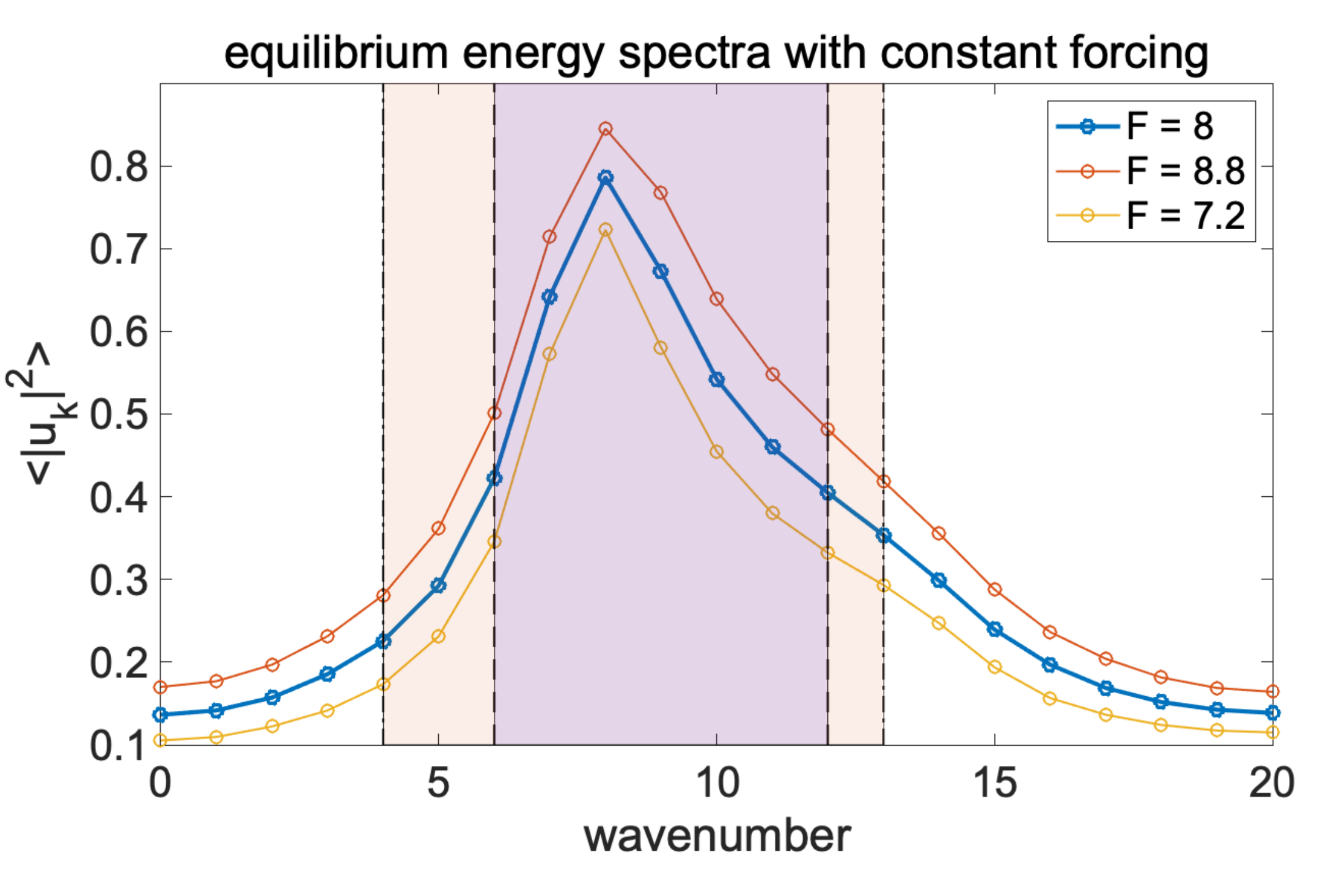}\includegraphics[scale=0.3]{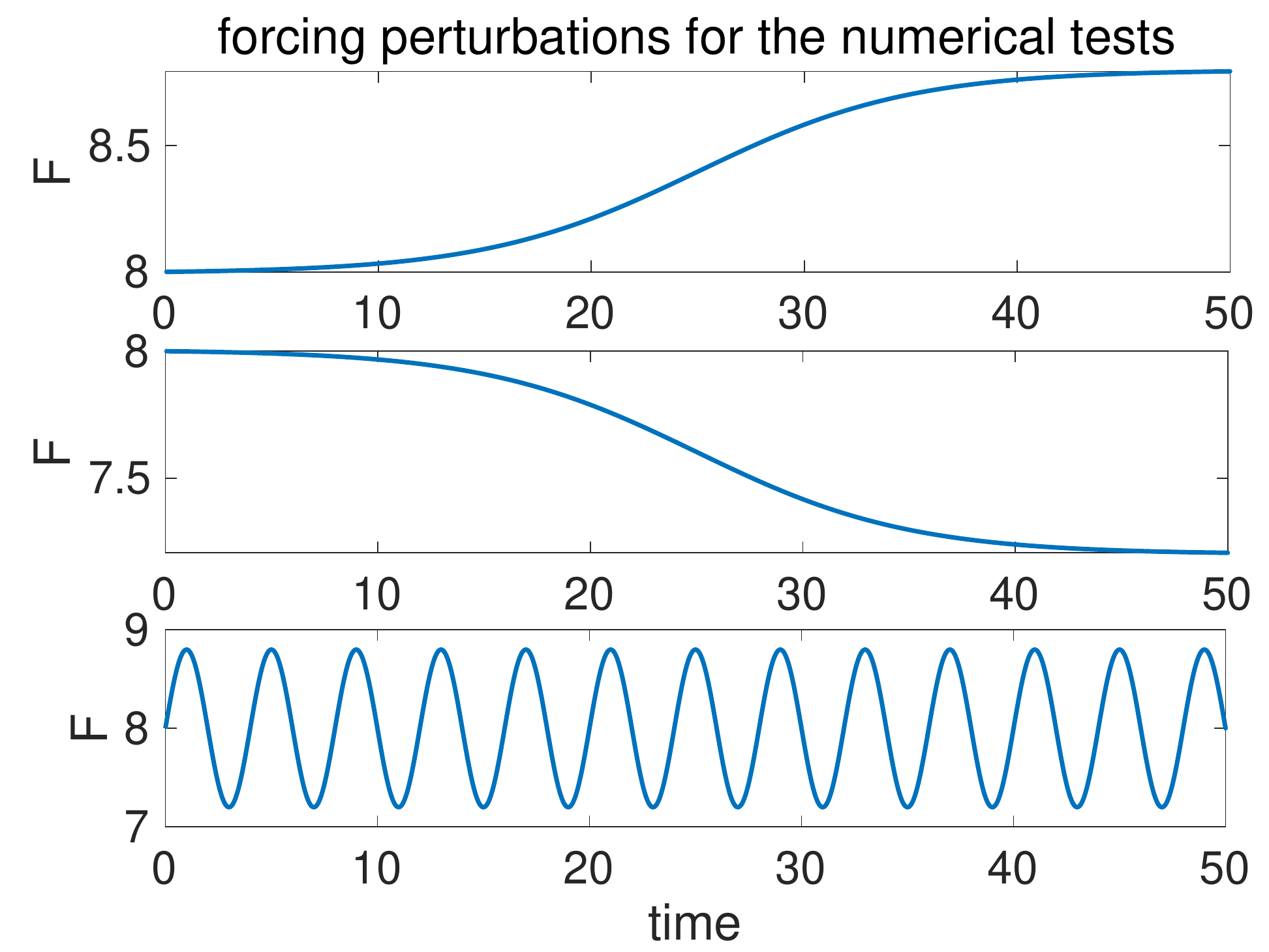}
\end{center}
\caption{Direct Monte-Carlo simulation solutions of the 40-mode L-96 system as the standard test model. The
left panel shows the equilibrium energy spectra and the inner shaded area
includes the resolved modes in the reduced-order model (\ref{discretenonmarkov2}). Unstable modes span
in a wider range $4\protect\leq k\protect\leq13$ than
the resolved state in the reduced-order model. The right panel shows several external forces we will consider for testing the prediction skill.}
\label{fig:Direct-Monte-Carlo-simulation}
\end{figure}

For simplicity, we consider uniform damping and forcing terms, $d\left(t\right)$ and $F\left(t\right)$ respectively, that are only functions
of time and identical for any grid points $j=0,\ldots J-1$. For an extensive test of the model prediction skill, we will consider several forcing functions with distinctive features (see the right panel in Figure~\ref{fig:Direct-Monte-Carlo-simulation}). With this assumption, the first two moments can be further simplified to a uniform mean state, $\bar{\mathbf{u}}\left(t\right)=\bar{u}\left(t\right)\left(1,\cdots,1\right)^{\mathrm{T}}$, and a diagonal covariance matrix, $R\left(t\right)=\mathrm{diag}\left(r_{0}\left(t\right),...,r_{J/2}\left(t\right)\right)$.
The corresponding moment equations are given as,
\addtocounter{equation}{0}\label{eq:stat_eqn}
\begin{subequations}
\begin{eqnarray}
\frac{d\bar{u}\left(t\right)}{dt} &= &  -d\left(t\right)\bar{u}\left(t\right)+\phi\left(t\right)+F\left(t\right),\label{eq:mean_L96}\\
\frac{dr_{k}\left(t\right)}{dt} &= &  -2\left[\Gamma_{k}\bar{u}\left(t\right)+d\left(t\right)\right]r_{k}\left(t\right)+\theta_k\left(t\right),\quad k=0,1,...,J/2,\label{eq:cov_L96} \\
\frac{dE\left(t\right)}{dt} &= & -2d\left(t\right)E\left(t\right)+F\left(t\right)\bar{u}\left(t\right),\label{eq:energy_eqn}
\end{eqnarray}
\end{subequations}
where we have defined the coupling coefficients $\Gamma_{k}=\frac{1}{J}\left(\cos\frac{4\pi k}{J}-\cos\frac{2\pi k}{J}\right)$,
$r_{-k}=\left\langle Z_{-k}Z_{-k}^{*}\right\rangle =\left\langle Z_{k}Z_{k}^{*}\right\rangle =r_{k}$,
the variance feedback $\phi$ to the mean equation, and the nonlinear flux $\theta_{k}$ in the variance equations,
\[
\phi=\sum_{k}r_{k}\Gamma_{k},\quad\theta_{k}=2\sum_{m}\mathfrak{Re}\left\{ \left\langle Z_{m}Z^{*}_{m+k}Z_{k}\right\rangle \left(e^{-2\pi i \frac{2m+k}{J}}-e^{2\pi i \frac{m+2k}{J}}\right)\right\},
\]
respectively, with statistical energy conservation $\mathrm{tr}(\theta_k)=0$. See Appendix A in \cite{majda2016improving} for a detail derivation of these terms.
A numerical discretization of the right hand sides of (\ref{eq:mean_L96})-(\ref{eq:energy_eqn}) gives an explicit example for the abstract operators $\mathcal{F}_1, \mathcal{F}_2, \mathcal{F}_3$ in (\ref{discretenonmarkov}).

\subsection{Mean-covariance closure model}

Here, we will specify the closure model for the discrete dynamical system in the form \eqref{discretenonmarkov} induced by the time discretization of \eqref{eq:mean_L96}-\eqref{eq:energy_eqn}. In Section~\ref{sec4}, we will numerically validate the effectiveness of the machine learning strategy in recovering the dynamical maps $\mathcal{G}_k$ that model the evolution of the nonlinear flux $\theta_k$ that are missing in this formulation,
\begin{equation}
\begin{aligned}\label{discretenonmarkov1}
\bar{u}_{i+1} &= \mathcal{F}_1(\bar{u}_{i},\{r_{k,i}\}_{k=0,\ldots,J/2},F_{i+1}),\\
r_{k,i+1} &= \mathcal{F}_{2,k}(\bar{u}_{i},r_{k,i},\theta_{k,i}), \quad\quad\quad\quad, \\
E_{i+1} &= \mathcal{F}_3(\bar{u}_{i},E_{i},F_{i+1}) \\
\theta_{k,i+1} &=\mathcal{G}_k (\bar{u}_{i},\ldots,\bar{u}_{i-m+1};\{\theta_{k,i},\ldots,\theta_{k,i-m+1}\}_{k=0,\ldots,J/2}; E_i,\ldots, E_{i-m+1})
\end{aligned}
\end{equation}
for $k = 0,\ldots, J/2$. Here, we should point out that $\bar{u}_i,{r}_{k,i},E_{i},\theta_{k,i}$ are all real-valued scalar variables, where we used subscript-$i$ to denote the discrete time index. Here, we have adopted another simplification by ignoring the explicit dependence of $\mathcal{G}_k$ on $\{r_{k}\}$. This simplification is partly motivated by the implicit dependence of the variance information through $E$.
Numerically, this simplification avoids the complexity in training the neural network model in approximating $\mathcal{G}:=(\mathcal{G}_0,\ldots,\mathcal{G}_{J/2})$. In such a case, we should point out that $\{\mathcal{G}_k:\mathbb{R}^{(3+J/2)m}\to\mathbb{R}\}_{k=0,\ldots,J/2}$ is already high-dimensional when $m$ is large, even without the explicit dependence on $\{r_k\}$. 

\subsection{Reduced-order mean-covariance closure model}
Next, we will consider a reduced-order model by truncating the summation term in \eqref{eq:mean_L96} to only account for leading modes 
$\mathcal{K}:=\{k \in \mathbb{Z}: k_{\min}\leq k\leq k_{\max}\}$ that carry large variances (see the variances as functions of modes on left panel of Figure \ref{fig:Direct-Monte-Carlo-simulation} for various constant forcings). In our numerical experiment, we consider $k_{\min}=6$ and $k_{\max}=12$ such that the resolved subset $\mathcal{K}$ includes only the most unstable modes .

Specifically, the reduced-order model is given by a coupled system consisting of 
\[
\frac{d\bar{u}\left(t\right)}{dt} =  -d\left(t\right)\bar{u}\left(t\right)+\sum_{k\in\mathcal{K}}r_{k}\left(t\right)\Gamma_{k}+\tilde{\phi}(t)+ F(t), 
\]
and the dynamics of $\{r_k\,:\, k \in \mathcal{K}\}$ in \eqref{eq:cov_L96} and the dynamics of $E$ in \eqref{eq:energy_eqn}. The key idea here is to consider a non-Markovian model (to be learned via appropriate machine learning algorithm) for the evolution of the unresolved total variance feedback $\tilde{\phi}:=\sum_{k\in\{0,\ldots,J/2\}\backslash\mathcal{K}}r_{k}\left(t\right)\Gamma_{k}$. In discrete form, our task will be to learn the dynamical maps $\mathcal{G}_{k,1}$ and $\mathcal{G}_2$ of,
\begin{equation}
\begin{aligned}\label{discretenonmarkov2}
\bar{u}_{i+1} &= \mathcal{F}_1(\bar{u}_{i},\{r_{k,i}\}_{k\in\mathcal{K}},F_{i+1})+ \tilde{\phi}_i,\\
r_{k,i+1} &= \mathcal{F}_{k,2}(\bar{u}_{i},r_{k,i},\theta_{k,i}), \\
E_{i+1} &= \mathcal{F}_3(\bar{u}_{i},E_{i},F_{i+1}), \\
\theta_{k,i+1} &=\mathcal{G}_{1,k} (\bar{u}_{i},\ldots,\bar{u}_{i-m+1};\{\theta_{k,i},\ldots,\theta_{k,i-m+1}\}_{k\in\mathcal{K}}; E_i,\ldots, E_{i-m+1}),\\
\tilde{\phi}_{i+1} &=\mathcal{G}_2 (\bar{u}_{i},\ldots,\bar{u}_{i-m+1}; \tilde{\phi}_i,\ldots, \tilde{\phi}_{i-m+1}; E_i,\ldots, E_{i-m+1}),
\end{aligned}
\end{equation}
for $k\in\mathcal{K}$ using appropriate machine learning algorithms. With the reduced-order model \eqref{discretenonmarkov2}, we only need to learn $\{\mathcal{G}_{1}:\mathbb{R}^{(2+|\mathcal{K}|)m}\to\mathbb{R}^{|\mathcal{K}|},\mathcal{G}_2:\mathbb{R}^{3m}\to\mathbb{R}\}_{k\in\mathcal{K}}$, where $\mathcal{G}_1:=(\mathcal{G}_{1,k})_{k\in\mathcal{K}}$ and we have denoted the number of modes in $\mathcal{K}$ by $|\mathcal{K}|=k_{\max}-k_{\min}+1$. Compare to the full-order closure model in \eqref{discretenonmarkov1}, $\mathcal{G}_1$ is a lower-dimensional map which makes the computational cost less expensive when $|\mathcal{K}|<J/2$.
In our numerical simulation, we will consider $\mathcal{K} = \{6,\ldots,12\}$ such that $|\mathcal{K}|=7< 21$ and the closure map to be recovered, $\mathcal{G}_1:\mathbb{R}^{9m}\to \mathbb{R}^7$, is a much smaller dimensional map relative to that in the full-order model where $\mathcal{G}_1:\mathbb{R}^{24m}\to \mathbb{R}^{21}$.

\subsection{Mean closure model}
Finally, we will consider a closure model that ignores the detail evolution of the covariance terms $r_k$. In such a severe truncation scenario, we will introduce a non-Markovian closure for $\phi := \sum_{k=-J/2+1}^{J/2}r_{k}\left(t\right)\Gamma_{k}$ to account for the combined contribution of the truncated covariance terms in \eqref{eq:mean_L96}. The corresponding discrete form is given as follows,
\begin{equation}
\begin{aligned}\label{discretenonmarkov3}
\bar{u}_{i+1} &= \mathcal{F}_1(\bar{u}_{i},F_{i+1})+ \phi_i,\\
E_{i+1} &= \mathcal{F}_3(\bar{u}_{i},E_{i},F_{i+1}), \\
\phi_{i+1} &=\mathcal{G} (\bar{u}_{i},\ldots,\bar{u}_{i-m+1}; \phi_i,\ldots, \phi_{i-m+1}; E_i,\ldots, E_{i-m+1}).
\end{aligned}
\end{equation}
Computationally, we only need to learn one map $\mathcal{G}:\mathbb{R}^{3m}\to\mathbb{R}$, which is a significant reduction compared to the previous models in \eqref{discretenonmarkov1} and \eqref{discretenonmarkov2}.

\subsection{An important strategy for modeling unstable dynamics}\label{sec:flux_decomp}

In the full-order and reduced-order covariance models \eqref{discretenonmarkov1} and \eqref{discretenonmarkov2}, neural network models will be constructed to update the variances $r_{k}$. One major challenge is the inclusion of strong inherent instability that
is common among turbulent dynamical systems. For example, in the L-96
system, the covariance equation (\ref{eq:cov_L96}) for $r_{k}$ contains positive
unstable modes with positive Lyapunov exponents if $-\Gamma_{k}\bar{u}>0$.
A lack of careful consideration in the detailed balance in unstable
variance dynamics will lead to unbounded model divergence in the numerical
verification. Particularly, an empirically trained neural-network model for the map $\mathcal{G}$ in \eqref{discretenonmarkov1} (or $\mathcal{G}_1$ in \eqref{discretenonmarkov2}) may not produce marginally stable dynamics that maintain accurate long-term stable forecasts.

To address this issue, we consider a more structural modeling, adopting the
ideas in \cite{majda2018strategies} by including an explicit nonlinear coupling
terms in the variance equation. To illustrate this, we modify the dynamical equation for $\mathcal{G}$ in \eqref{discretenonmarkov1} and \eqref{discretenonmarkov2} as follows:
We decompose the higher-order nonlinear flux $\theta_{k}$ containing all the third moments
in a (nonlinear) effective damping $d_{k,i+1}^{M}$ and noise $\sigma_{k,i+1}^{M}$ 
such that 
\begin{equation}
\begin{aligned}\theta_{k,i+1} & =-d_{k,i+1}^{M}r_{k,i}+\sigma_{k,i+1}^{M},\\
Q_{k,i+1}^{M} & =\mathcal{G}_{k}\left(\bar{u}_{i},\cdots,\bar{u}_{i-m+1};\{\theta_{k,i},\cdots,\theta_{k,i-m+1}\}_{k=0,\ldots,J/2};E_{i},\cdots,E_{i-m+1}\right)\\
d_{k,i+1}^{M} & =-\min\left\{ Q_{k,i+1}^{M},0\right\} /r_{k,\mathrm{eq}},\\
\sigma_{k,i+1}^{M} & =\max\left\{ Q_{k,i+1}^{M},0\right\} .
\end{aligned}
\label{eq:clos2}
\end{equation}
Here, the map $\mathcal{G}_{k}$ models the full nonlinear flux at each time instant
and we employ an LSTM network in the next section to approximate $\mathcal{G}:=(\mathcal{G}_{0},\dots,\mathcal{G}_{J/2})$. 
However, instead of directly setting $\theta_{k,i+1}=Q_{k,i+1}^{M}$, which gives the last equation in \eqref{discretenonmarkov1}, we split the model output into two
positive effective damping $d_{k}^{M}>0$ and effective noise $\sigma_{k}^{M}>0$.
The effective damping is recovered from the unperturbed equilibrium
statistics $r_{k,\mathrm{eq}}$. In this way, the unstable directions in the system are stabilized by the effective damping modeling the nonlinear transfer of energy without altering the detailed statistical balance in the equilibrium. One can see that if $Q_{k,i+1}^M$ is positive (that is, the mode is stable), then
$\theta_{k,i+1}=Q_{k,i+1}^M$ and we retain the original model in \eqref{discretenonmarkov1}.

\section{Machine learning of the missing Non-Markovian components}\label{sec3}

In this section, we briefly discuss how to employ the Long-Short-Term-Memory (LSTM) \cite{Hochreiter_1997}, a recurrent neural-network, to learn the hidden non-Markovian maps in the proposed closure statistical models in \eqref{discretenonmarkov1}, \eqref{discretenonmarkov2}, \eqref{discretenonmarkov3}. To simplify the discussion, let us identify the input variable (or covariate) with a sequence of  
correlated state variables $\left\{ \mathbf{x}_{j}\right\} _{j=i-m+1}^{i}$ measured at $m$ time instants ahead of the prediction time $i+1$ and the output (response) variable at discrete time index-$i+1$ as $\mathbf{y}_{i+1}$. In the case of \eqref{discretenonmarkov1}, the input variable is $\mathbf{x}_j =\{\bar{u}_j,\theta_{0,j},\ldots,\theta_{J/2,j}, E_j\}$ and the output variable is $\mathbf{y}_{i+1}=\{\theta_{0,i+1},\ldots,\theta_{J/2,i+1}\}$. For \eqref{discretenonmarkov2}, the input variable is $\mathbf{x}_j =\{\bar{u}_j,\{\theta_{k,j}\}_{k\in\mathcal{K}},\tilde{\phi}_j, E_j\}$ and the output variable is $\mathbf{y}_{i+1}=\{\{\theta_{k,i+1}\}_{k\in\mathcal{K}},\tilde{\phi}_{i+1}\}$. For \eqref{discretenonmarkov3}, the input variable is $\mathbf{x}_j =\{\bar{u}_j,\phi_j, E_j\}$ and the output variable is $\mathbf{y}_{i+1}=\{\phi_{i+1}\}$.

Recurrent neural networks offer the desirable structure to incorporate temporal processes of sequential data with long temporal correlations and keep tracking of hidden processes. The LSTM network is designed to avoid the problem of vanishing gradients. The building block of LSTM is to consider the following model, which is known as an LSTM cell,
\begin{equation}
\begin{aligned}\mathbf{f}_{i}= & \sigma_{g}\left(W_{f}\mathbf{x}_{i}+U_{f}\mathbf{h}_{i-1}+V_{f}\mathbf{c}_{i-1}+\mathbf{b}_{f}\right),\\
\mathbf{I}_{i}= & \sigma_{g}\left(W_{i}\mathbf{x}_{i}+U_{i}\mathbf{h}_{i-1}+V_{i}\mathbf{c}_{i-1}+\mathbf{b}_{i}\right),\\
\mathbf{c}_{i}= & \mathbf{f}_{i}\otimes \mathbf{c}_{i-1}+\mathbf{I}_{i}\otimes\tanh\left(W_{c}\mathbf{x}_{i}+U_{c}\mathbf{h}_{i-1}+\mathbf{b}_{c}\right),\\
\mathbf{o}_{i}= & \sigma_{g}\left(W_{o}\mathbf{x}_{i}+U_{o}\mathbf{h}_{i-1}+V_{o}\mathbf{c}_{i}+\mathbf{b}_{o}\right),\\
\mathbf{h}_{i}= & \mathbf{o}_{i}\otimes\tanh\left(\mathbf{c}_{i}\right).
\end{aligned}
\label{eq:lstm}
\end{equation}
In \eqref{eq:lstm}, $\sigma_{g}=\frac{1}{1+e^{-x}}$ is the sigmoid activation
function, and $\otimes$ represents the element-wise product. The
model cell includes forget, input, and output gates $\mathbf{f}_{i},\mathbf{I}_{i},\mathbf{o}_{i}$,
and the cell state $\mathbf{c}_{i}$. The hidden process $\left\{ \mathbf{h}_{i-m+1},\cdots,\mathbf{h}_{i-1},\mathbf{h}_{i}\right\} $
represents the time-series of the unresolved process. In a compact form, let us denote the LSTM cell in \eqref{eq:lstm} as $\mathbf{h}_{i+1}=\mathrm{Lc}\left(\mathbf{x}_{i},\mathbf{h}_{i}\right)$, 
where we have suppressed the dependence on the parameters for simplicity.

The LSTM network is constructed from $m$ LSTM cells $\mathrm{Lc}$
with the same structure and parameters $\mathbf{W}$. The cells are
connected by the intermediate hidden state $\mathbf{h}_{i}\in\mathbb{R}^{h}$.
Every LSTM cell takes in the input data $\mathbf{x}_{i}$ at the $i$-th
step and the output $\mathbf{h}_{i}$ from the previous adjacent cell,
and gives out the inner hidden state $\mathbf{h}_{i+1}$ to be used
for prediction of the next state. The full LSTM chain is connected
through $m$ sequential cell structures, that is,
\begin{equation}
\mathbf{h}_{m}=\mathrm{Lc}^{\left(m\right)}\left\{ \mathbf{h}_{0};\mathbf{x}_{i-m+1},\cdots,\mathbf{x}_{i}\right\} \equiv\mathrm{Lc}\left(\mathbf{x}_{i}\right)\circ\cdots\circ\mathrm{Lc}\left(\mathbf{x}_{i-m+1}\right)\left(\mathbf{h}_{0}\right),\label{eq:lstm_full}
\end{equation}
where the composition operator is defined with respect to the hidden state $\mathbf{h}_i$. In \eqref{eq:lstm_full}, the data at different time instance, $\mathbf{x}_{i}$, is fed
into the corresponding LSTM cell, and the hidden state $\mathbf{h}_{i}$ is the output of the previous cell and input for the next cell.
For simplicity, the initial value of the hidden state is often set
as zero, $\mathbf{h}_{0}=0$. The final output $\mathbf{h}_{m}$ from
the last step of the LSTM chain goes through a final single layer
fully connected linear model given as,
\begin{equation}
\hat{\mathbf{y}}_{i+1}=A\mathbf{h}_{m}+\mathbf{b},\label{eq:lstm_final}
\end{equation}
where $A\in\mathbb{R}^{d_y\times h},\mathbf{b}\in\mathbb{R}^{d_y}$ are the model coefficients in the final layer and $d_y=\mathrm{dim}(\mathbf{y})$ denotes the dimension of the output variables. In our numerical implementation, for the reduced-order model in \eqref{discretenonmarkov2}, we consider two LSTM networks, one for estimating $\mathcal{G}_1$ and another one for estimating $\mathcal{G}_2$. One can also consider separate LSTM networks for each component $\mathcal{G}_k$ in \eqref{discretenonmarkov1} (or $\mathcal{G}_{1,k}$ in \eqref{discretenonmarkov2}), which we do not pursue in our numerical experiments.

\subsection{Empirical loss functions}
The neural network parameters $\mathbf{W}:=\{A,\mathbf{b},W_f,U_f,V_f,\mathbf{b}_f,W_i,U_i,V_i,\mathbf{b}_i,\ldots\}$ are obtained by solving a nonlinear non-convex optimization problem to minimize the difference between the training output data $\{\mathbf{y}_{j}^\ell\}_{\ell,j=1}^{n,M}$ and the LSTM output data 
$\{\hat{\mathbf{y}}_{j}^{\ell}\}_{\ell,j=1}^{n,M}$, subjected to the same input data $\{\mathbf{x}_j^\ell\}_{j=i-m+1,\ell=1}^{i,n}$, where $n$ denotes the total number of training samples. There are many ways to design loss functions. Denote the true output data as $\mathbf{y}_{j}^{\ell} =\{\phi_{j}^{\ell},\theta_{k,j}^{\ell}\}$ and the LSTM model output data as $\hat{\mathbf{y}}_{j}^{\ell} = \{\hat{\phi}_{j}^{\ell},\hat{\theta}_{k,j}^{\ell}\}$, then we can, for example, consider the following empirical loss function,
\BEA
\sum_{j=1}^{M}\left[\alpha\sum_{\ell=1}^n  \big(\phi^{\ell}_{j}-\hat{\phi}_{j}^{\ell}(\mathbf{W}^\phi)\big)^2 +\sum_{k}\beta_{k}\sum_{\ell=1}^n \big|\theta_{k,j}^\ell-\hat{\theta}_{k,j}^{\ell}(\mathbf{W}^\theta)\big|\right], \label{eq:loss_flux}
\EEA
where we have defined $\mathbf{W}^\phi$ and $\mathbf{W}^\theta$ to distinguish the parameters of the two network models. For the full-order mean-covariance model in \eqref{discretenonmarkov1}, we set $\alpha=0$ and minimize \eqref{eq:loss_flux} for $\mathbf{W}^\theta$. For the reduced-order mean-covariance model in \eqref{discretenonmarkov2}, we set $\alpha=1$ and $\beta_k>0$ such that they have comparable scales and
minimize \eqref{eq:loss_flux} for both $\mathbf{W}^\phi$ and $\mathbf{W}^\theta$.
 For the mean closure model in \eqref{discretenonmarkov3}, we set $\alpha=1$ and $\beta_k=0, \forall k$, and minimize \eqref{eq:loss_flux} for $\mathbf{W}^\phi$. While other choices exist, such as to include the error in the mean and variance components, we found the improvement is not significant. We should also point out that the empirical loss function in \eqref{eq:loss_flux} is defined over a path of length-$M$, the model parameters are obtained by one minimization problem. In practice, we found that with $M=10$, the resulting estimate yields more stable long-time predictions, compared to just setting $M=1$ (for which one can solve separate minimization problems to obtain independent LSTM networks for $\mathcal{G}_1$ and $\mathcal{G}_2$, by fitting to one-step forecast data as employed in \cite{HJLY:19}). 
 
\subsection{Small training dataset}

\begin{figure}
\begin{center}
\includegraphics[width=.82\textwidth]{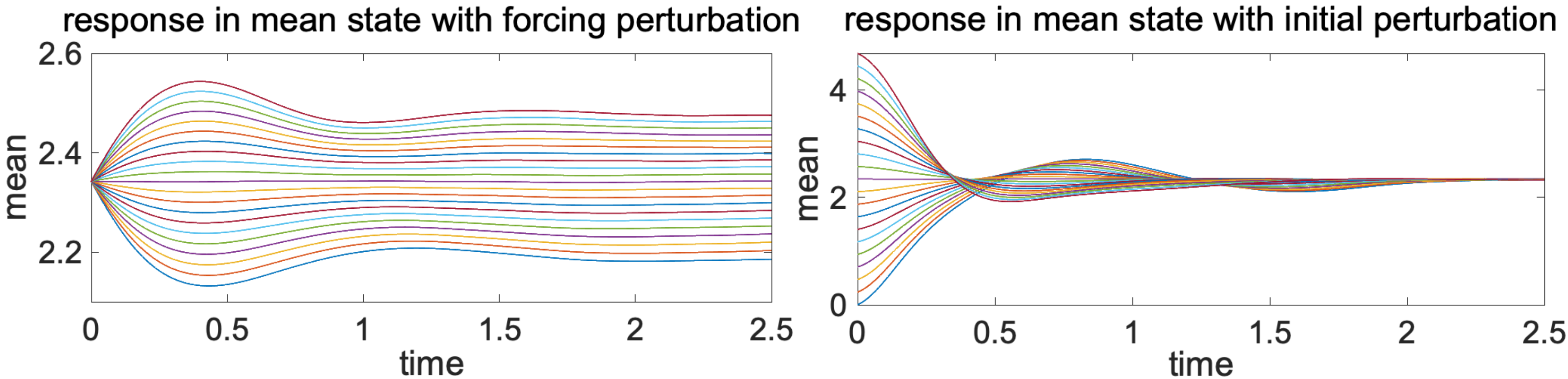} \\
{\normalsize (a) Response in $\bar{u}$} 
\end{center}
\vspace{-2em}
\begin{center}
\includegraphics[width=.82\textwidth]{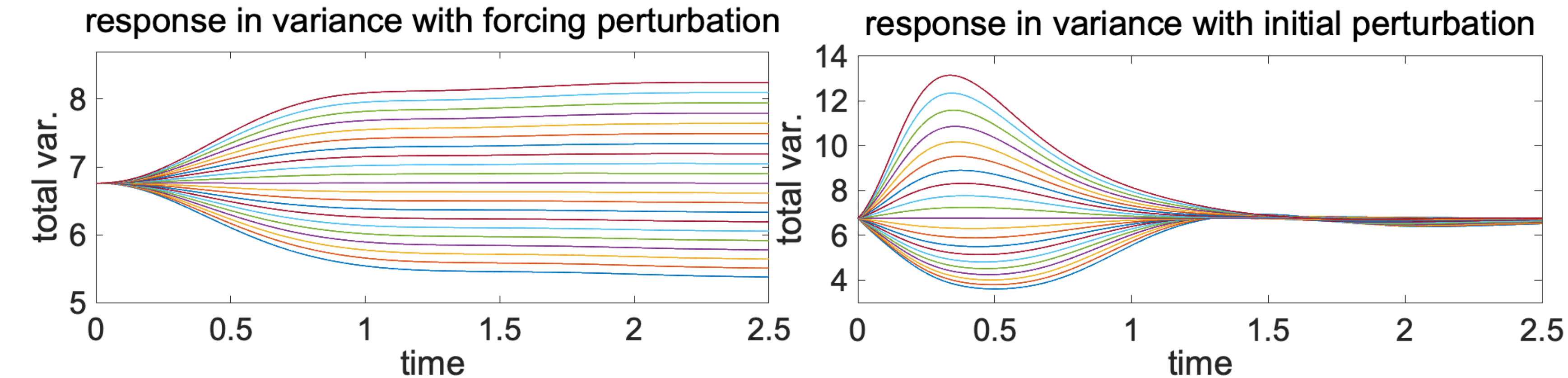} \\
{\normalsize (b) Response in $\mathrm{tr}(R)$}
\end{center} 
\caption{Statistical responses correspond to perturbations in external forcing
 (left) and in the initial mean state (right). Responses
of the statistical mean and total variance with different perturbation
amplitudes are shown.}
\label{fig:Responses}
\end{figure}

While the general unperturbed underlying non-Markovian dynamics in \eqref{discretenonmarkov} is an example of the missing dynamical model formulated in \cite{jh:20,HJLY:19},
the statistical configuration here is more challenging due to the shortage of informative training data that reflect the key features of the underlying dynamical process. In our numerical test problem (the L-96 model), the statistics are homogeneous such that the statistics of each of the solutions forced by a constant forcing will decay to a constant value in a short time (see Figure~\ref{fig:Responses}). To compensate for the lack of observed statistical data, in practice, we generate the training data by a direct short-time Monte-Carlo simulation following these steps:
\begin{itemize}
\item[i.] Generate an ensemble of unperturbed equilibrium statistical solutions with the reference forcing $F=F_{\mathrm{eq}}$. Each ensemble member solves an initial value problem corresponding to a randomly drawn initial condition from the standard Gaussian distribution.

\item[ii.] Simulate an ensemble of solutions to the statistical steady state subjected to various constant external perturbations $F=F_{\mathrm{eq}}+\delta f$. The ensemble of solutions at the final time from [i.] is used as initial conditions. The empirical mean and variance of these initial conditions correspond to the mean state  $\bar{u}=\bar{u}_{\mathrel{eq}}\approx 2.35$, and total variance,
$\mathrm{tr}(R) = \mathrm{tr}(R_{eq})\approx 6.8$, respectively (see the left panel of Figure~\ref{fig:Responses}). 

\item[iii.] Simulate an ensemble of solutions to the statistical steady state correspond to unperturbed constant external perturbation $F=F_{\mathrm{eq}}$ and perturbed initial conditions $\bar{u}\rightarrow\bar{u}+\delta\bar{u}$. We perturb each ensemble member of the initial condition by adding a constant value $\delta\bar{u}$ to each ensemble member at the final time from [i.]. In the right panel of Figure~\ref{fig:Responses}, one can see that the ensemble mean states at the initial time vary while the total variances at the initial time stay the same.  
\end{itemize}

Typical statistical trajectories with different initial and forcing perturbation amplitudes are depicted in Figure~\ref{fig:Responses}. The statistics of the perturbed states, which exhibit strong nonlinear coupling effects, decay to new (or original unperturbed) equilibrium states beyond the decorrelation time. Notice that for this problem, the decaying behavior of these trajectories yields a small training dataset (in terms of temporal length). Beyond the transient time, the time series saturates and, thus, is not informative.

In general, even when long time statistics are informative (e.g., for nontrivial time-dependent external forces or non-stationary statistical dynamics), from a practical standpoint, attaining longer time series is computationally infeasible, especially when the dimension of the underlying state space is high and/or the system requires a stiff numerical solver. Thus the configuration that we consider (training with an ensemble of short time series) in the present paper can be used for a wide class of high-dimensional systems when long time series are not accessible. 

\section{Numerical results}\label{sec4}

We now examine the effectiveness of the statistical model schemes discussed above with detailed numerical tests. We start with the full-order mean-covariance model \eqref{discretenonmarkov1} to learn the unresolved high-order nonlinear flux directly from data. Second, the reduced-order mean-covariance model \eqref{discretenonmarkov2} is proposed for efficient computation with the most energetic leading modes. Finally, we show an even more efficient computation of the mean statistical prediction using only the mean closure model \eqref{discretenonmarkov3}, focusing on the mean responses subjected to various forcing perturbations. 

\subsection{Model configuration for training and prediction}
In the training stage, the training data are generated from $41$ response solutions (shown in Figure~\ref{fig:Responses}) with either perturbed initial states $\delta\bar{u}\in\left[-\bar{u}_{\mathrm{eq}},\bar{u}_{\mathrm{eq}}\right]$ or constant forcing perturbations $\delta f\in\left[-0.1F_{\mathrm{eq}}, 0.1F_{\mathrm{eq}}\right]$ from direct Monte-Carlo
solutions of the L-96 system with an ensemble size $10000$. The true equation (\ref{eq:L96}) is integrated
with a 4th-order Runge-Kutta scheme with a small time step $\delta t=0.001$, while the data is sampled at every 10 steps. Thus we have the data sampling step $\Delta t=0.01$. The training model is updated $M=10$ times to account for the integrated error along the time integration. Notice that this choice of larger measurement step size leads to numerical discretization errors in computing the time integration and recovering the parameters of $\phi$ and $\theta_{k}$. 
The total number of samples is $n=1640$, and they are obtained by collecting non-overlapping time interval $M\Delta t=0.1$ units from the statistical response trajectories. With such a small sample size, the learning problem is rather challenging as the neural-network model has a large number of parameters. While one can, of course, generate more data by additional perturbations and initial conditions, we will not pursue this direction since our goal is to understand the effectiveness of the agnostic machine learning model in such a stringent configuration with a small training dataset.

In the prediction stage, we verify the model
performance by considering the long-time statistical prediction under a variety of time-dependent forcing scenarios that are not observed in the training dataset. For long-time prediction, the model output in the previous step is reiterated as an input in the next forecast stage, thus model errors accumulate in time. Therefore, it requires the closure models to be numerically stable in resistance to the accumulated model errors in the neural network model. In our numerical tests, we consider the ramp-type forcings and the periodic forcing as standard test examples where the large external perturbation is introduced to its equilibrium state forcing, $F_{\mathrm{eq}}=8$ (see Figure \ref{fig:Direct-Monte-Carlo-simulation} for changes in the energy spectra for different forcing perturbations). In application, such testing configurations can be used to simulate the climate change scenario where the original state is driven away from its previous equilibrium state due to external perturbations \cite{majda2016improving,majda2019linear} and other uncertainty quantification tasks.

In addition, a residual
structure is adopted in the neural network for the closure models (\ref{discretenonmarkov1})-(\ref{discretenonmarkov3})
\BEA
\theta_{i+1}=\mathcal{G}=\theta_{i}+\tilde{\mathcal{G}},\label{resnet}
\EEA
where $\tilde{\mathcal{G}}$ denotes the LSTM network \eqref{eq:lstm_full} to update the increment of the unresolved higher-order component. The LSTM chain contains $m=100$ repeating cells with the same structure,
taking a time sequence of time length $T=1$ which is still shorter than the correlation time of the system (see Figure \ref{fig:Responses}).
The dimensions of the hidden states in LSTM are taken as $h_{v}=50$ for the variance equation and $h_{m}=10$ for the mean equation.
The optimization for the loss \eqref{eq:loss_flux} is carried out by the ADAM scheme. A total of 100 epochs is repeated during training, starting from the learning rate $\mathrm{lr}=5\times10^{-4}$
which is reduced three times to half of its original value at the epoch
number 25, 50, and 75.

\subsection{Prediction skill of the mean-covariance model}\label{sec4.1}

In this section, we numerically verify the prediction skill of the full statistical mean-covariance
model (\ref{discretenonmarkov1}). For clarity, we split the discussion into two subsections. First, we state the concrete discrete closure model corresponding to this example. Subsequently, we report the detailed prediction skill. 

\subsubsection{Training model to learn the unresolved nonlinear flux}

In the full mean-covariance model in \eqref{eq:mean_L96}-\eqref{eq:energy_eqn}, the
dynamical equations for the mean state $\bar{u}$, the total energy $E$, and variance $r_k$ are given explicitly and  Markovian. 
In our numerical experiment, the discrete form in \eqref{discretenonmarkov1} is obtained by adopting the mid-point implicit scheme on
\eqref{eq:mean_L96}-\eqref{eq:energy_eqn} to ensure a more robust numerical performance with the larger time step $\Delta t=10\delta t$. Together with the 
structural form in \eqref{eq:clos2} to avoid instabilities and the residual network architecture in \eqref{resnet}, the overall dynamical closure model adopted here is given as follows:
\begin{equation}
\begin{aligned}\bar{u}_{i+1}-\bar{u}_{i} & =\Delta t\left[-\frac{d}{2}\left(\bar{u}_{i}+\bar{u}_{i+1}\right)+\sum_{k}\frac{\Gamma_{k}}{2}\left(r_{k,i}+r_{k,i+1}\right)+\frac{1}{2}\left(F_{i}+F_{i+1}\right)\right],\\
E_{i+1}-E_{i} & =\Delta t\left[-d\left(E_{i}+E_{i+1}\right)+\frac{1}{2}\left(\bar{u}_{i}F_{i}+\bar{u}_{i+1}F_{i+1}\right)\right],\\
r_{k,i+1}-r_{k,i} & =\Delta t\left[-\Gamma_{k}\left(\bar{u}_{i}r_{k,i}+\bar{u}_{i+1}r_{k,i+1}\right)-d\left(r_{k,i}+r_{k,i+1}\right)+\theta_{k,i+1}\right],\\
\theta_{k,i+1} - \theta_{k,i} & =\min\left\{ Q_{k,i+1}^{M},0\right\} /r_{k,\mathrm{eq}}r_{k,i}+\max\left\{ Q_{k,i+1}^{m},0\right\} ,\\
Q_{k,i+1}^{M} & =\mathcal{G}_{k}\left(\bar{u}_{i},\cdots,\bar{u}_{i-m+1},\{\theta_{k,i},\cdots,\theta_{k,i-m+1}\}_{k=0,\ldots,J/2},E_{i},\cdots,E_{i-m+1}\right) 
\end{aligned}
\label{eq:model1_nn}
\end{equation}
In \eqref{eq:model1_nn}, the states are discretized at time intervals $t_{i+1}-t_{i}=\Delta t$. The exact dynamical equations for the mean $\bar{u}$, the total statistical energy $E$, and $r_k$ are adopted, while dynamics of
$\theta_{k}$ are learned from data, with $\mathcal{G}_k$ modeled by an LSTM architecture. 

\subsubsection{Numerical results for detailed mean and variance prediction}

\begin{figure}
\subfloat[upward ramp forcing]{\includegraphics[scale=0.32]{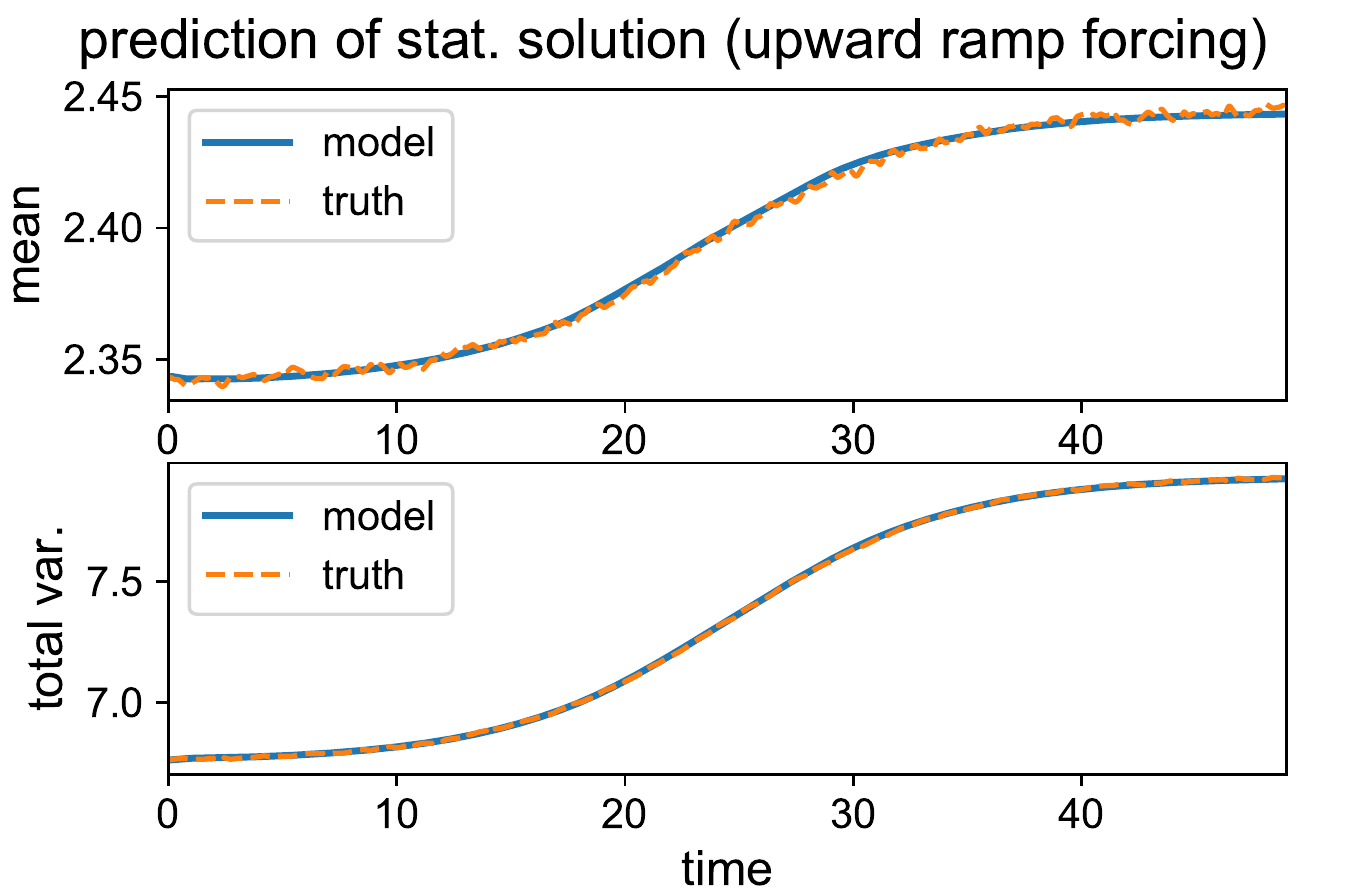}

}\subfloat[downward ramp forcing]{\includegraphics[scale=0.32]{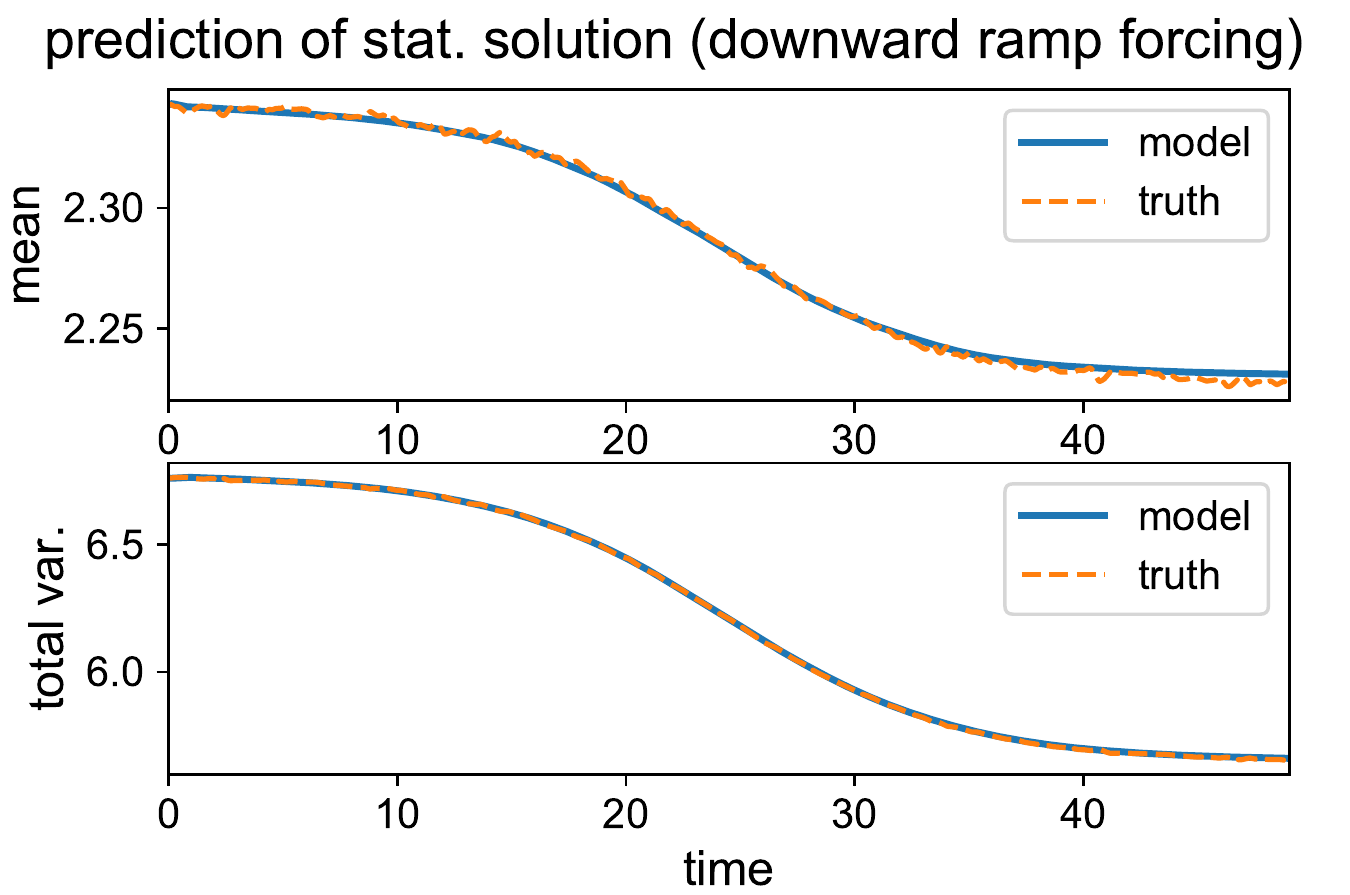}

}\subfloat[periodic forcing]{\includegraphics[scale=0.32]{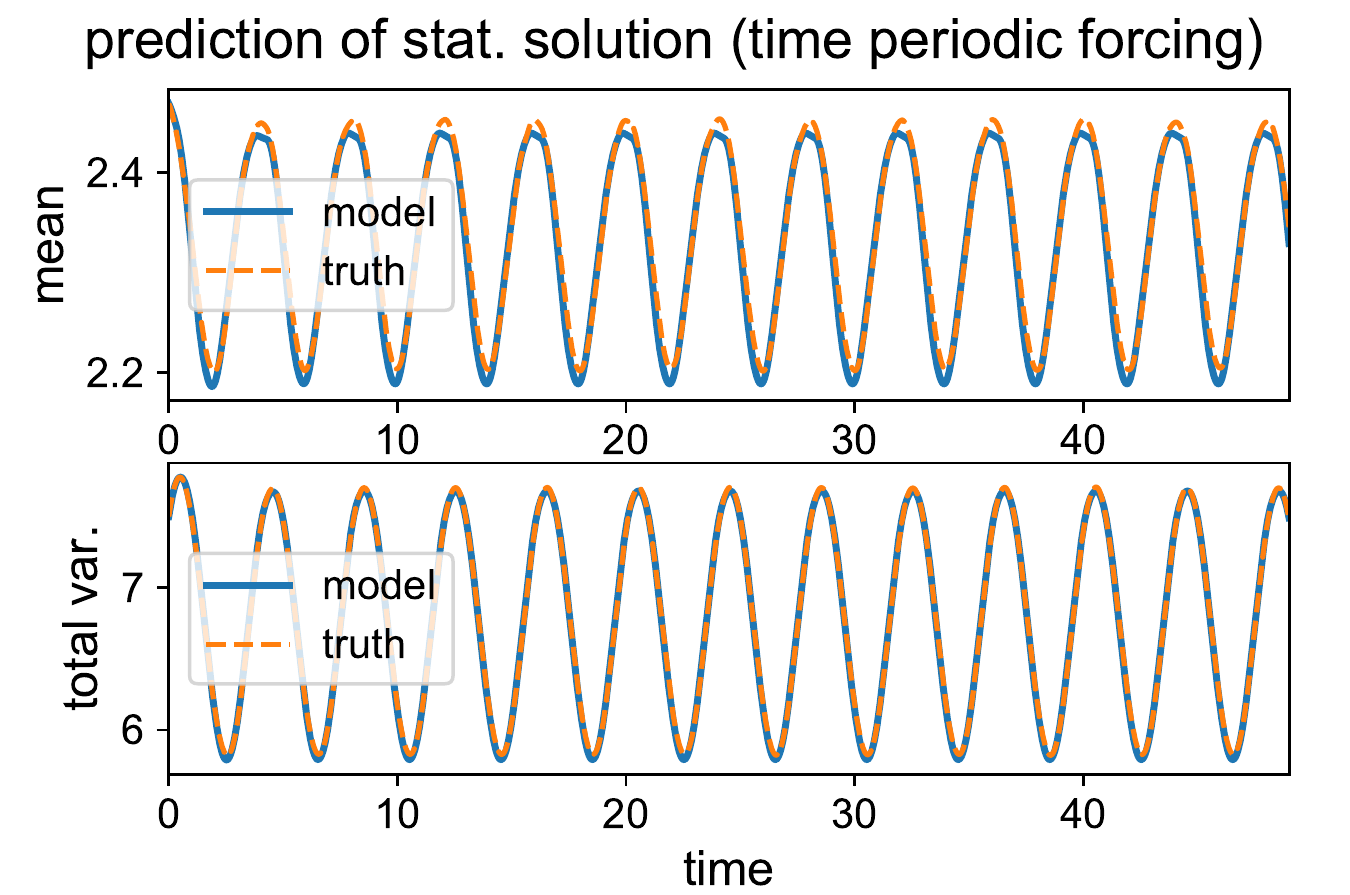}

}

\caption{Long-time model prediction with the full mean-covariance closure model. Predictions of the statistical mean and total variance under three different external forcing scenarios are compared.\label{fig:Long-time-model-prediction}}

\end{figure}

\begin{figure}
\begin{center}
\subfloat{\includegraphics[scale=0.43]{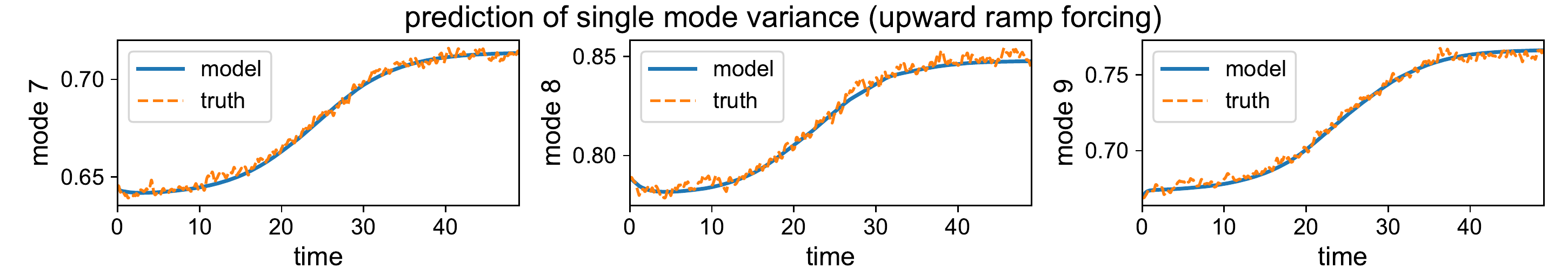}}
\vspace{-1.5em}
\subfloat{\includegraphics[scale=0.43]{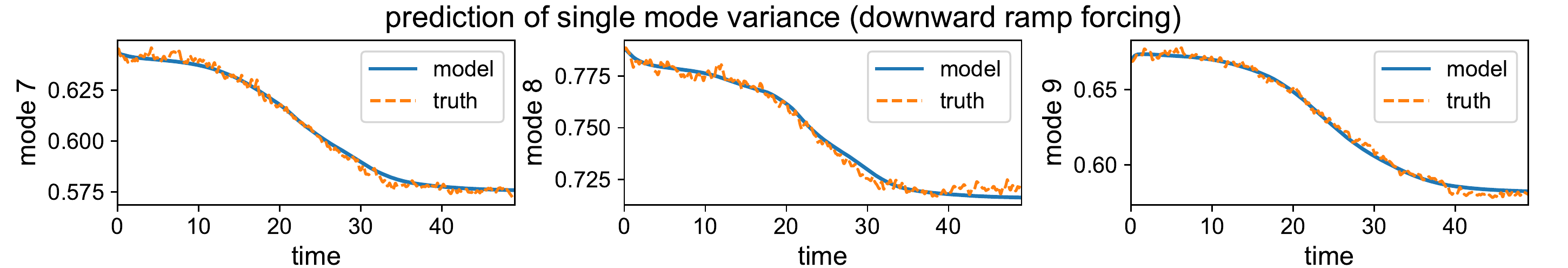}}
\vspace{-1.5em}
\subfloat{\includegraphics[scale=0.43]{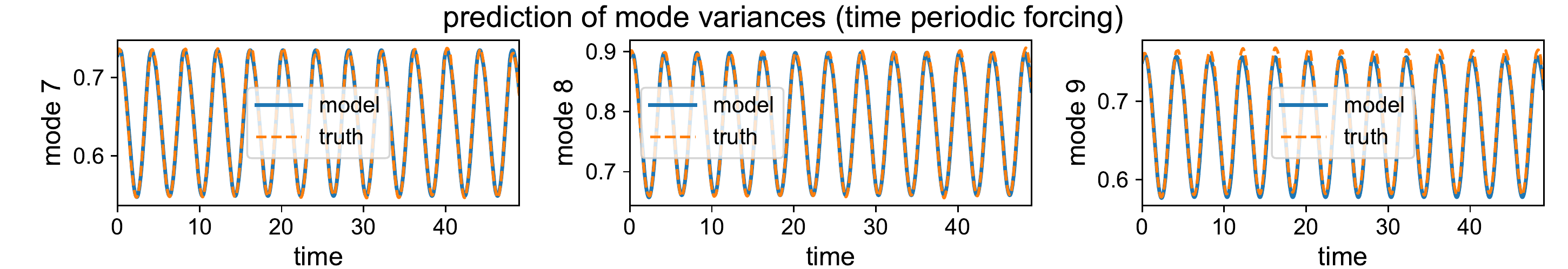}}
\end{center}
\caption{Detailed prediction of the variances of the first three most energetic modes
using the full mean-covariance model.\label{fig:Detailed-prediction-var}}
\end{figure}

Figure \ref{fig:Long-time-model-prediction} shows the model prediction
performance under the three forcing scenarios. The numerical model (\ref{eq:model1_nn}) is trained with a very short time dataset under
constant forcings (in Figure \ref{fig:Responses}), while the prediction performance is tested on time-dependent forcing perturbations (in the right panel of Figure \ref{fig:Direct-Monte-Carlo-simulation}). It
is shown that for the long-time prediction (up to $T=50$), the trained neural network model is stable and generates accurate predictions of both the statistical mean and variance throughout the time interval
among all three test cases. For a more detailed comparison of the variance response on individual mode, Figure \ref{fig:Detailed-prediction-var} compares the predictions of the variances of the first three leading modes. Again, we observe robust accurate prediction of variances in all the modes under the tested forcing cases containing different statistical features.


\begin{figure}
\centering
\includegraphics[scale=0.5]{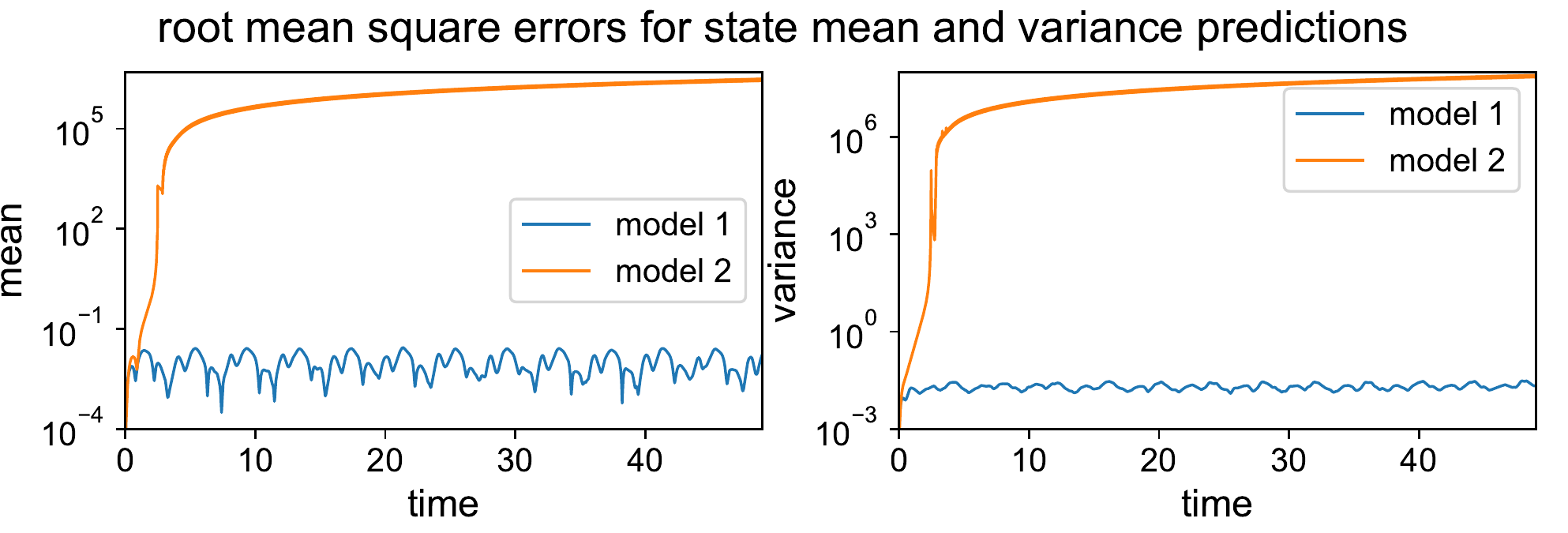}

\caption{Prediction RMSEs using the trained models with different model structures.
Model 1: the full-order model with decomposed effective damping
and noise in \eqref{eq:clos2}; Model 2: the direct model without using the proper nonlinear flux decomposition.\label{fig:Prediction-errors}}
\end{figure}

In addition, we confirm the importance of adopting the strategy discussed in Section \ref{sec:flux_decomp} to guarantee long-time numerical stability. The variance dynamics include
a large number of unstable directions that will amplify even small
errors. Considering this, the neural network approximation for $\theta_{k}$ adopts the decomposed structure (\ref{eq:clos2}) so that the marginally stable modes are balanced. Otherwise, if a neural network
is applied directly to the model structure $\theta_{k}$ without proper consideration of the physical mechanism, severe
numerical instability may occur due to the insufficient modeling of
the unstable dynamics. Numerically, we compare the root mean square errors (RMSEs) in mean and total variance prediction in Figure \ref{fig:Prediction-errors}. Indeed, we see that the optimal model with decomposed damping and noise structure
(model 1) maintains high accuracy for the long prediction period.
In contrast, if the nonlinear flux $\theta_{k}$ is directly learned
from the neural network (model 2), the predicted solution diverges
after a short time due to the strong inherent persistent instability 
in the system.

\subsection{Prediction skill of the reduced-order mean-covariance model}\label{sec4.2}

Next, we consider the reduced-order mean-covariance model for efficient
computation of only the most energetic modes $k_{\min}\leq k\leq k_{\max}$ in the variance equation. The total contribution of the less energetic unresolved modes is accounted with another neural network model for $\tilde{\phi}$. Thus
the computational scheme follows the discretized mean and variance
equations \eqref{discretenonmarkov2}, with the dynamical equation for $\theta$ being modified as in \eqref{eq:clos2} to avoid instability of the flux with residual network structure in \eqref{resnet}.

\comment{\begin{equation}
\begin{aligned}\bar{u}_{i+1}-\bar{u}_{i} & =\Delta t\left[-\frac{d}{2}\left(\bar{u}_{i}+\bar{u}_{i+1}\right)+\sum_{k_{\min}\leq k\leq k_{\max}}\frac{\Gamma_{k}}{2}\left(r_{k,i}+r_{k,i+1}\right)+\frac{1}{2}\left(F_{i}+F_{i+1}\right)+\phi_{i+1}\right],\\
r_{k,i+1}-r_{k,i} & =\Delta t\left[-\Gamma_{k}\left(\bar{u}_{i}r_{k,i}+\bar{u}_{i+1}r_{k,i+1}\right)-d\left(r_{k,i}+r_{k,I+1}\right)+\theta_{k,i+1}\right],\;k_{\min}\leq k\leq k_{\max},\\
E_{i+1}-E_{i} & =\Delta t\left[-d\left(E_{i}+E_{i+1}\right)+\frac{1}{2}\left(\bar{u}_{i}F_{i}+\bar{u}_{i+1}F_{i+1}\right)\right].
\end{aligned}
\label{eq:model2_nn}
\end{equation}
We introduce two neural networks for the unresolved mean feedback
$\phi_{i+1}$
and the higher-order flux term in the resolved variance equation $\theta_{k,n+1}$.
In this way, we are able to recover the mean state and the leading variance modes
without running the expensive full-order model. This is especially appealing to the high-dimensional  systems where the large number
of unresolved small scale fluctuation modes can be learned altogether
from data.
}

\begin{figure}
\begin{center}
\subfloat[upward ramp forcing]{\includegraphics[scale=0.32]{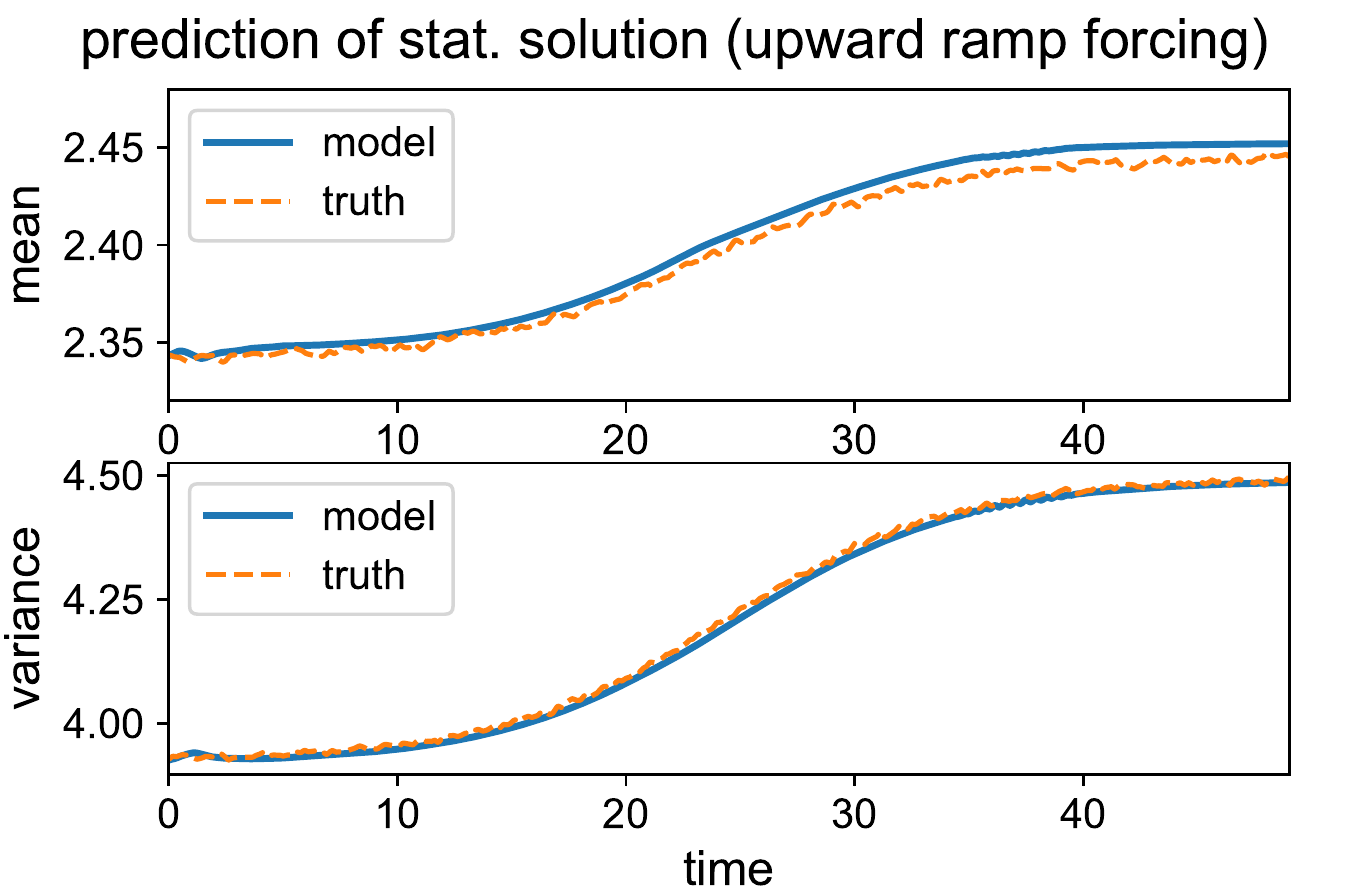}
}\subfloat[downward ramp forcing]{\includegraphics[scale=0.32]{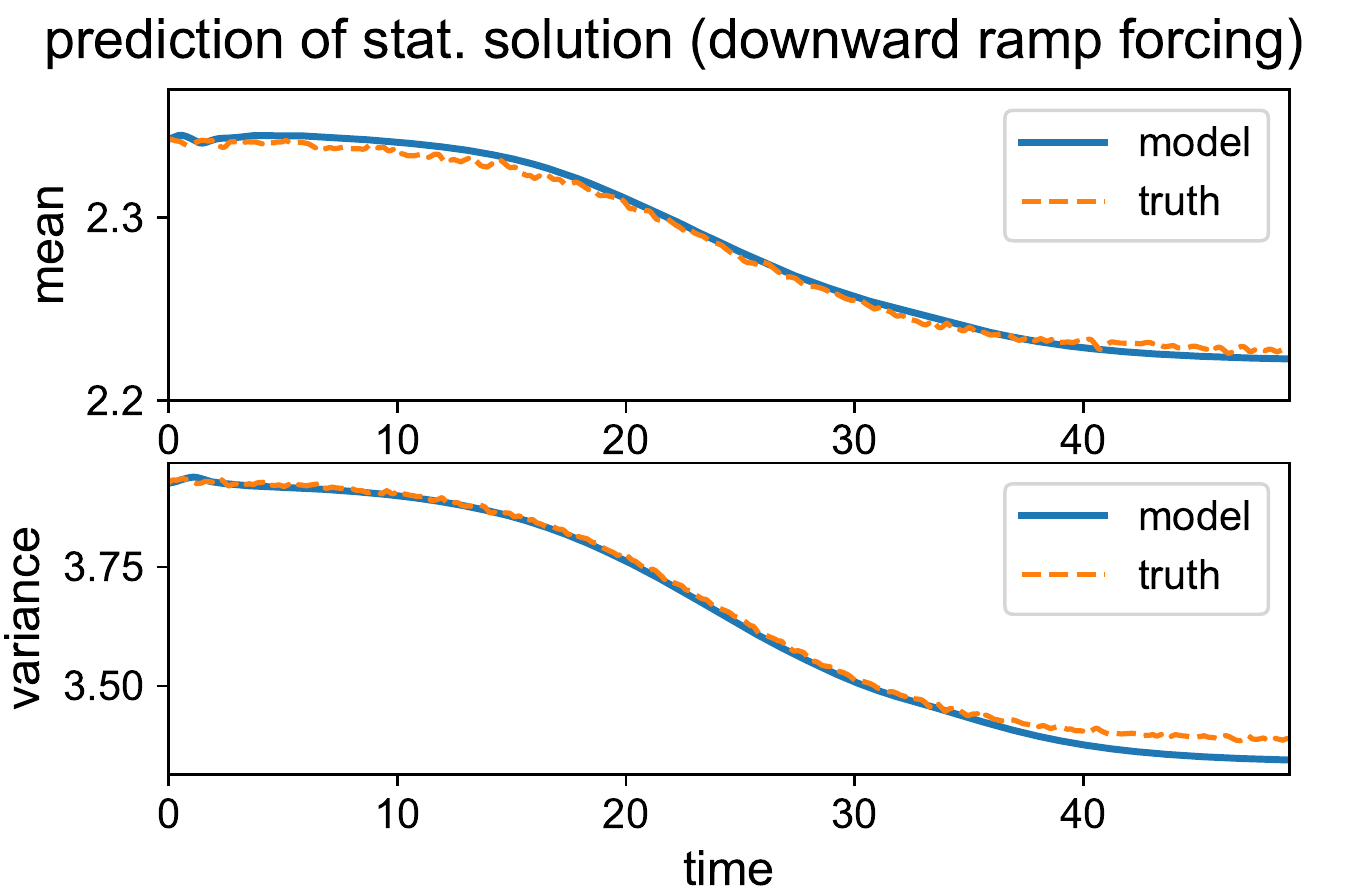}
}\subfloat[periodic forcing]{\includegraphics[scale=0.32]{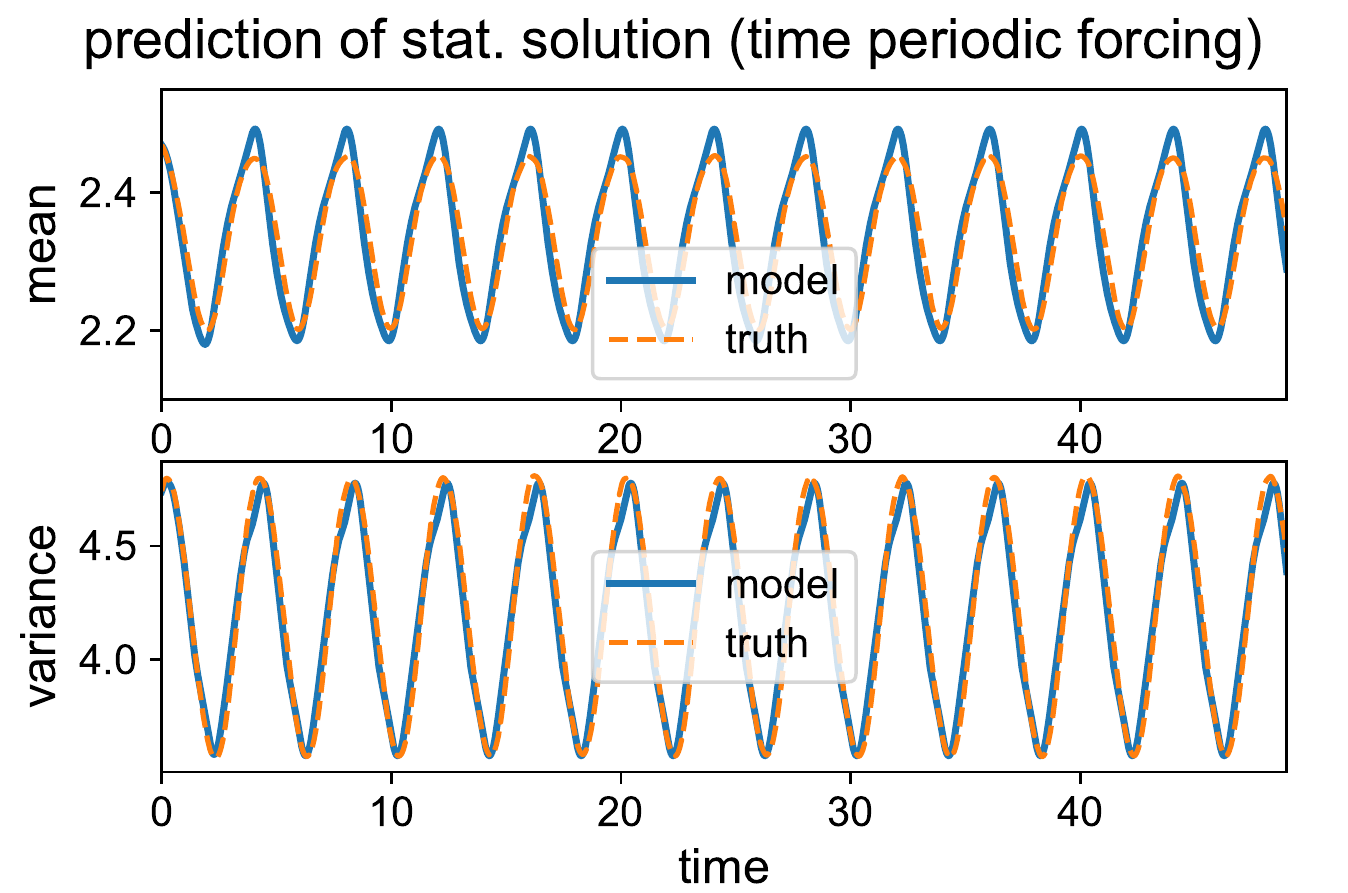}
}
\end{center}

\caption{Prediction of the statistical mean and variance on the resolved modes
using the reduced-order mean-variance model. The same neural network
model is applied to different types of external forcing forms. Only
the most energetic modes $6\protect\leq k\protect\leq12$ are computed
in the model.\label{fig:mvmodel-prediction}}
\end{figure}

In Figure \ref{fig:mvmodel-prediction}, we compare the reduced-order
model prediction for different forcing perturbations. Again,
we attain accurate predictions on both the statistical mean state and
variances in the resolved subspace among the different kinds of forcing cases. In comparison to the full-order model prediction in Figure~\ref{fig:Long-time-model-prediction},
a slightly larger error occurs here, especially in the mean state. This reflects the additional model error due to the model approximation for the many unresolved modes. However, the computation cost is significantly reduced since we only compute a small portion of the full system (7 out of the total 21 modes).

\begin{figure}
\begin{center}
\subfloat[machine learning model]{\includegraphics[scale=0.38]{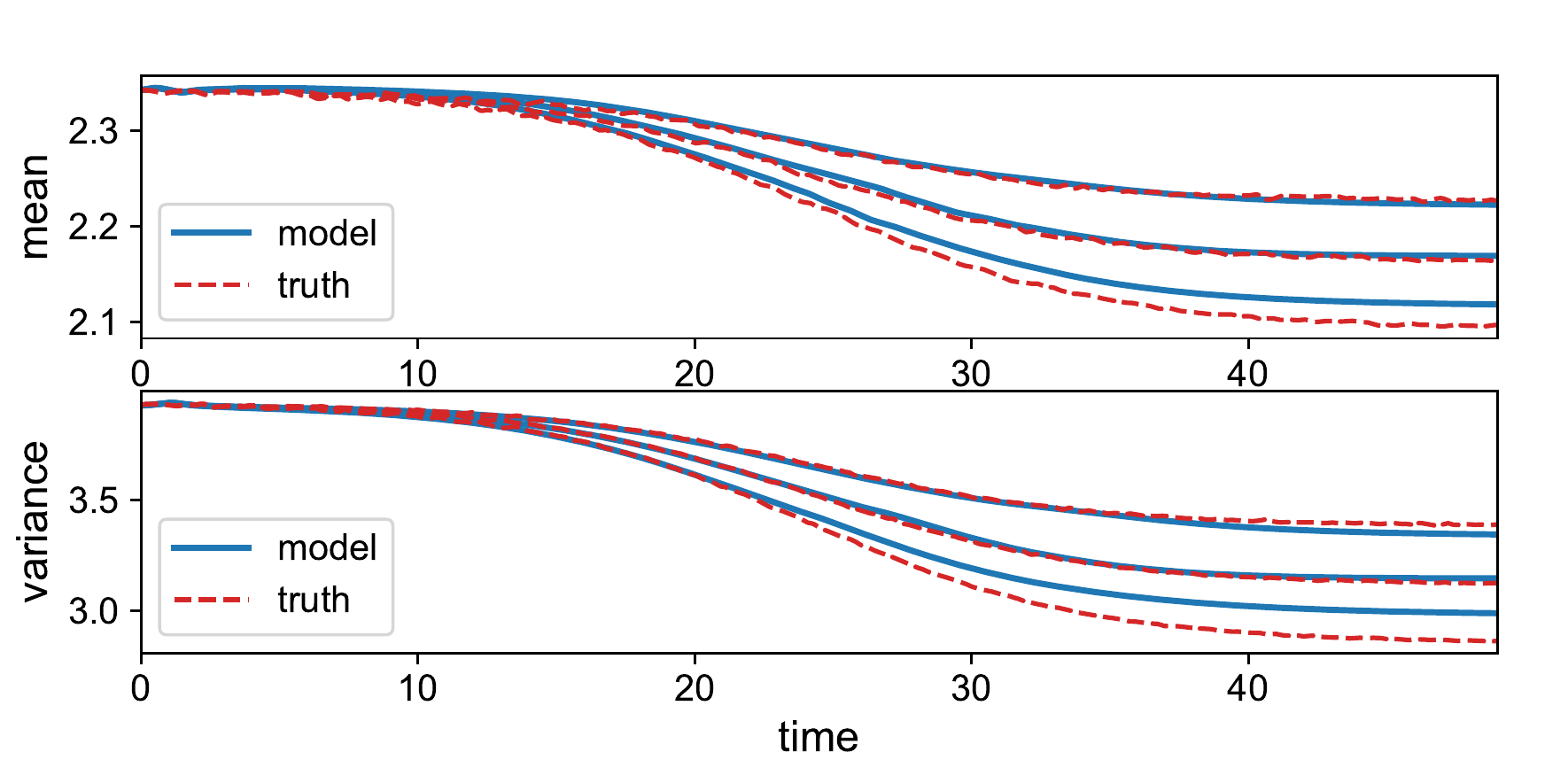}
}\hspace{-1em}\subfloat[parametric closure model]{\includegraphics[scale=0.28]{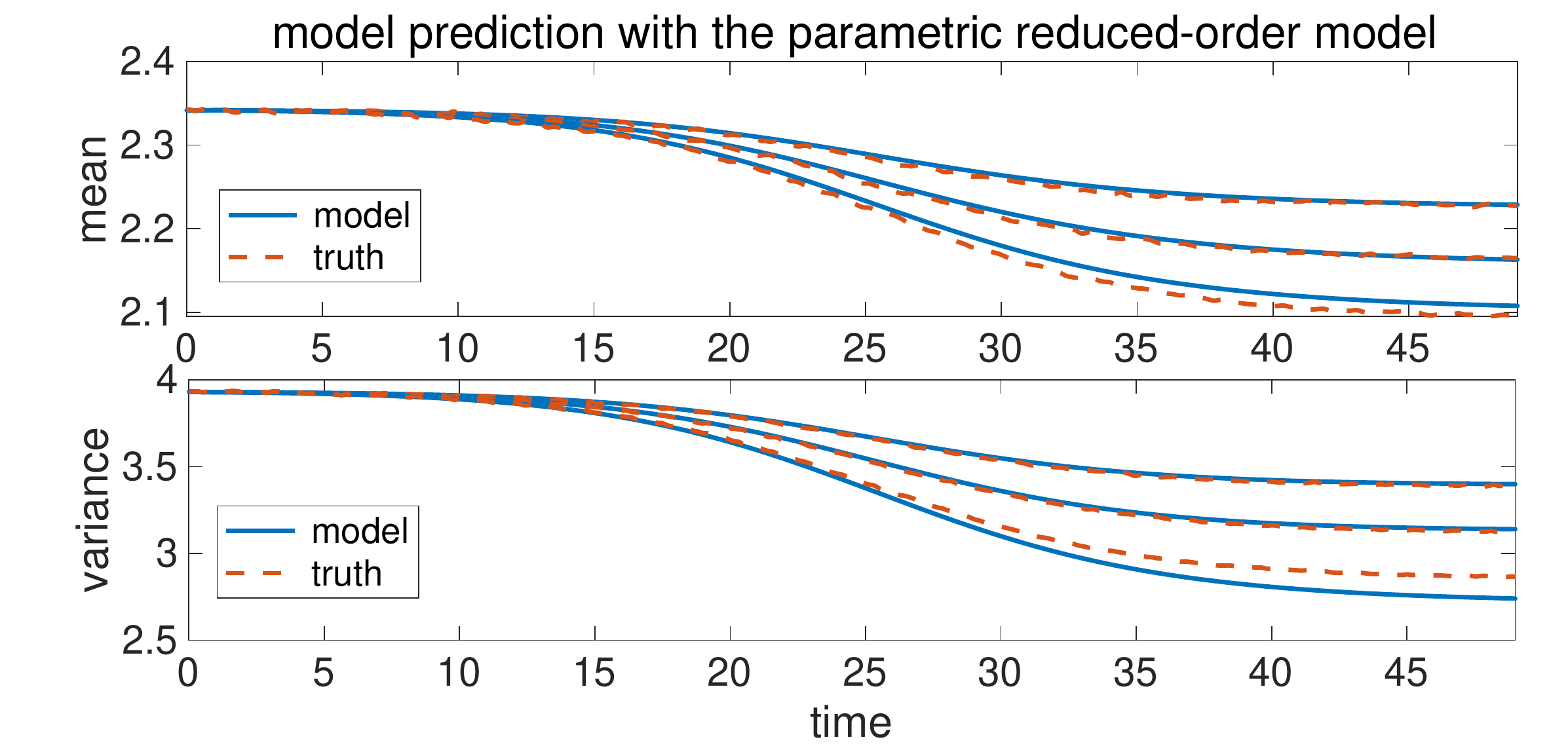}
}
\end{center}

\caption{Comparison of reduced-order model predictions under different forcing perturbation
amplitudes. The downward ramp forcing case is shown here as a typical example. The left panels show the predictions of machine learning model and the right panels show the prediction of the parametric closure model proposed in \cite{majda2016improving}. 
\label{fig:Comparison-of-model-amplitude}}
\end{figure}

Furthermore, to check the robustness of the model, we verify the model prediction skill with even stronger forcing perturbation amplitudes. Figure~\ref{fig:Comparison-of-model-amplitude} shows the downward forcing case with stronger maximum forcing perturbations, $\delta f=-0.1F_{\mathrm{eq}},-0.15F_{\mathrm{eq}},-0.2F_{\mathrm{eq}}$ (beyond the maximum forcing $|\delta f|=0.1F_{\mathrm{eq}}$ in the training data). Notice that the long-time prediction skill remains accurate for $-0.15F_{\mathrm{eq}}$ and starts to deteriorate for larger forcing amplitude, $-0.2F_{\mathrm{eq}}$. This somewhat negative result for larger perturbation is not so surprising as it displays the difficulty of the machine learning model in extrapolating beyond the information contents in the training data. 

As a benchmark, we also show the corresponding prediction skill of the parametric closure model \cite{majda2016improving} in the right panel of Figure \ref{fig:Comparison-of-model-amplitude}. By visual comparison, one can see that the prediction skill is very similar to the machine learning-based closure model; for the largest forcing amplitude, the parametric closure gives a slightly better prediction. 
While the prediction performance is comparable, this parametric model requires a complicated calibration strategy that involves an expensive brute-force minimization of a loss function that depends on a long time statistics (linear response statistics). Particularly, the evaluation of the loss function involves an integration of the reduced-order model for a long time for each choice of parameter. Beyond this step that can be expensive for high-dimensional problems (as the dimension of the parameter space increases), a more fundamental issue is that it requires a physical insight for choosing the parametric model for the flux term. 
On the other hand, the more agnostic neural-network model can capture the changes in the statistics without specifying some detailed nonlinear flux structure beyond \eqref{eq:clos2} that overcome instability.

\subsection{Prediction skill of the mean closure model}\label{sec4.3}

\begin{figure}
\centering
\includegraphics[scale=0.35]{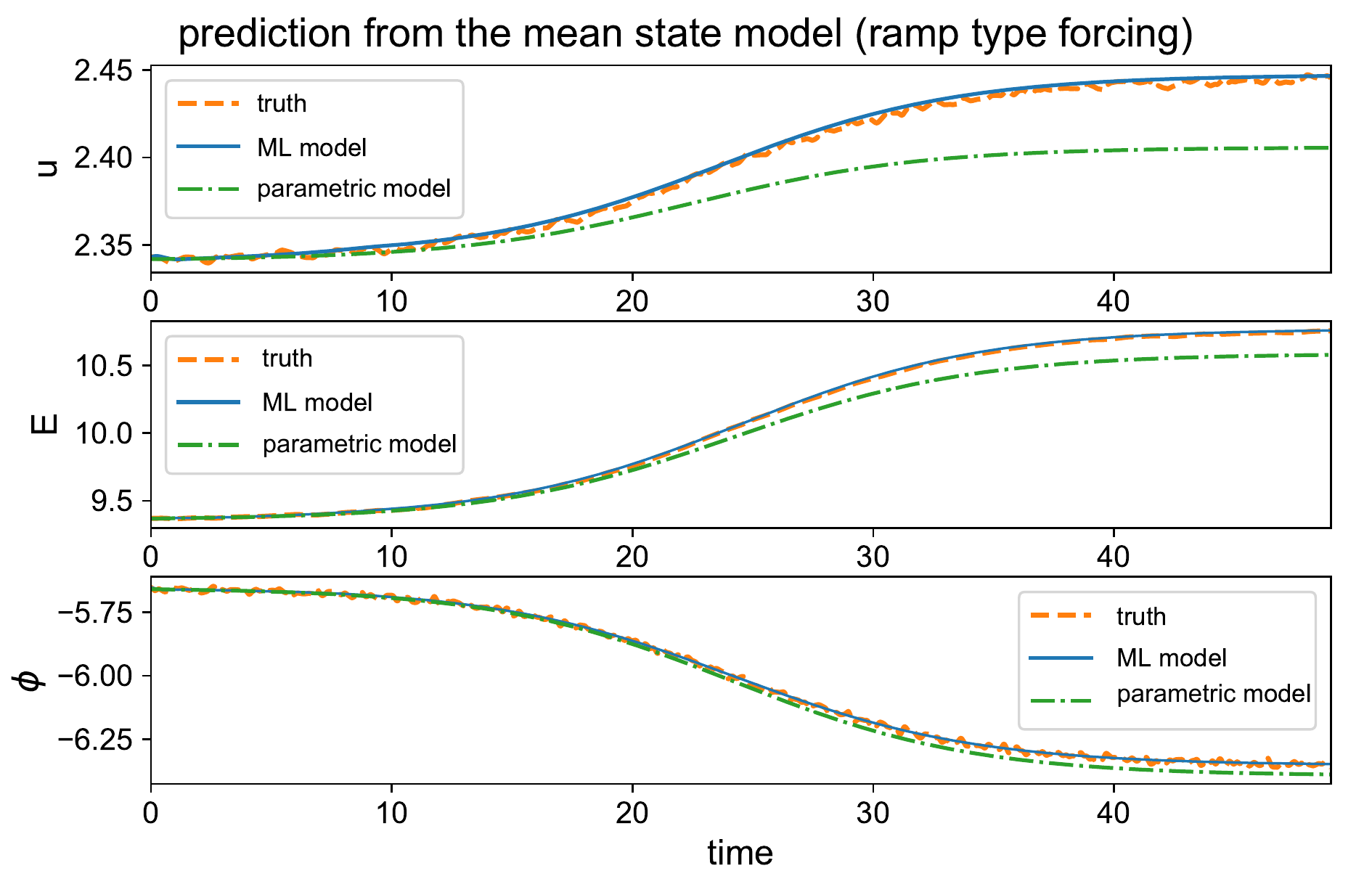}\includegraphics[scale=0.35]{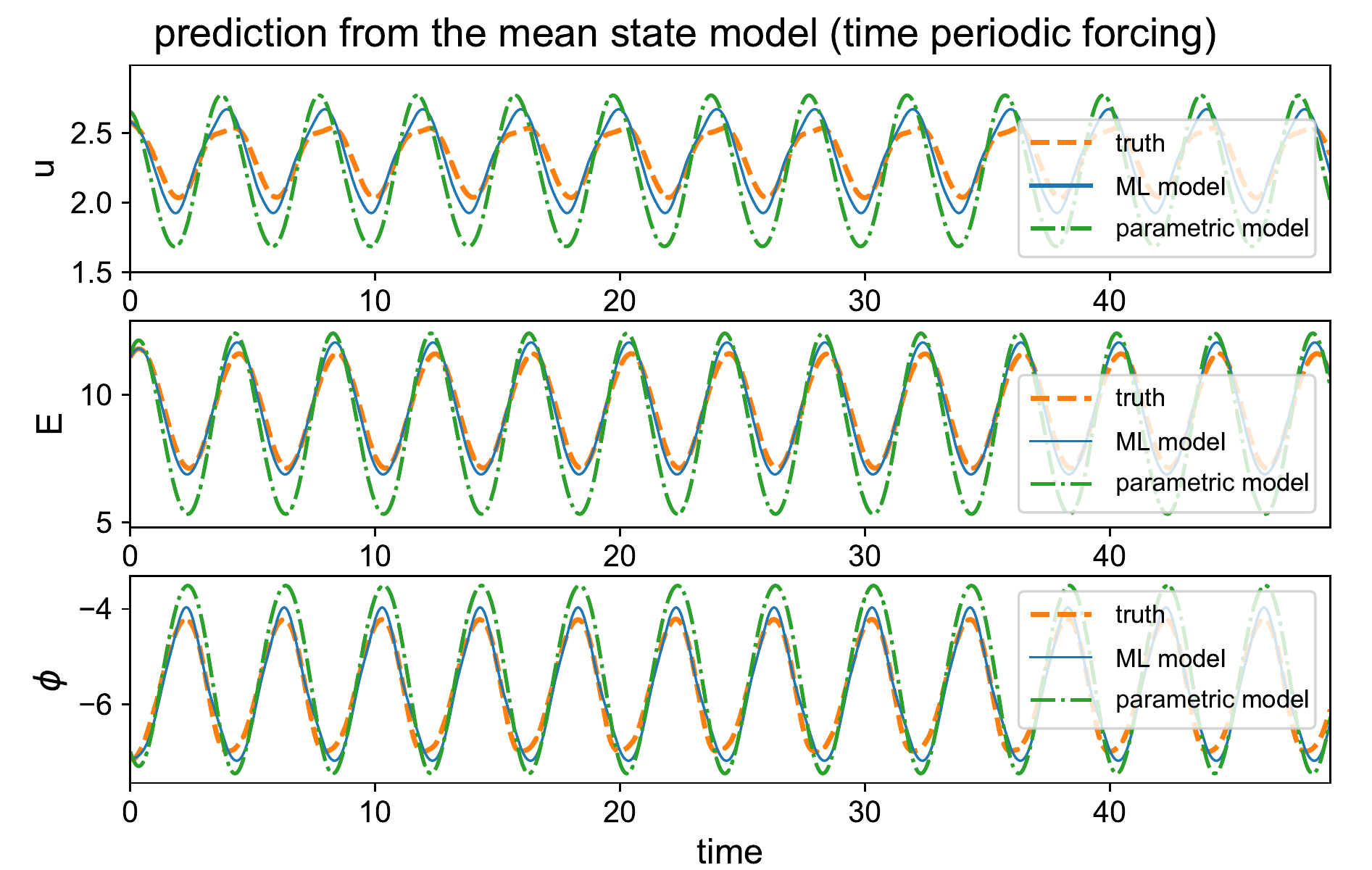}
\caption{Predictions for the mean state $\bar{u}$, total statistical energy
$E$, and the mean dynamical feedback $\phi$ with the mean statistical
model under different forcing scenarios. The optimized machine learning
model is compared with the parametric closure model.\label{fig:Long-term-prediction-mean}}

\end{figure}

Finally, we test the prediction skill of the mean closure model \eqref{discretenonmarkov3}, where we adopt the implicit midpoint rule for the discretization of \eqref{eq:mean_L96} and \eqref{eq:energy_eqn}, and use the standard LSTM network for the unresolved high-order
feedback. We test the performance of the neural network model for long-term
mean state prediction under the ramp down and periodic forcings in Figure~\ref{fig:Direct-Monte-Carlo-simulation}. In Figure \ref{fig:Long-term-prediction-mean}, we plot the predicted mean state $\bar{u}$, total statistical energy $E$, as well as
the variance feedback in the mean $\phi$. It shows that the machine learning model successfully
captures the changes in the mean state under this extreme model setup (with severely truncated dynamics) without including the explicit dynamical equations of the second-order moments. 
With forcing to a non-Gaussian regime (downward ramp forcing) or a periodic forcing with larger amplitude, the prediction becomes less accurate compared to the closure models that include more detailed variance dynamics (see e.g., Figures~\ref{fig:Long-time-model-prediction} and \ref{fig:mvmodel-prediction}). Still, the closure model maintains high prediction skills under the unseen forcings. Again, if we compare the machine learning model results with the parametric closure model in \cite{majda2016improving}, the machine learning framework produces more accurate predictions with cheaper computational costs. This shows the robustness of the agnostic neural network-based approach on various truncated configurations. On the other hand, the less accurate parametric model is due to the difficulty in modeling the truncated flux terms with a simple parametric equation. 

\section{Summary}\label{sec5}
In this paper, we developed non-Markovian statistical modeling strategies with machine learning.  In the construction of the statistical closure models, we considered learning the complicated dynamical structure of the high-order nonlinear flux terms directly from data by imposing the LSTM neural-network architecture to uncover the non-Markovianity induced by partially observed components. Three statistical mean and covariance models were considered, with different emphasis on the prediction of full variance spectrum, most energetic leading modes, and only the mean state. With limited training data due to the stationarity of the statistics beyond the correlation time, we enriched the training dataset by simulating the transient behavior of the statistics under various constant forcings and perturbed initial conditions.  

The performance of the hierarchical machine learning models was verified on the L-96 system with homogeneous statistics. Uniformly accurate long-time predictions are observed using the resulting ML model under different forcing perturbation functions and strong perturbation amplitudes beyond the data in the training set. In addition, the true nonlinear physical energy transfer mechanism was considered in the model construction to guarantee numerical stability in long-term numerical integration. The machine learning model displays strong resistance to accumulated model errors with a long-time stable prediction despite the inherent instability in a wide spectrum of modes in the L-96 system. We found that the ML-based model prediction is comparable to that of the existing parametric model, which requires a more detailed calibration strategy, in two scenarios: learning the full-order and reduced-order coupled systems of mean-covariance statistics. On the other hand, the ML model is more accurate than the parametric approach in the severely truncated regime, learning the closure of only the mean statistics.

From this study, we conclude that the agnostic machine learning model is portable on various truncation scenarios since the strategy does not require physical knowledge of the high-order flux terms (as in parametric modeling) beyond avoiding instabilities. Numerically, the proposed scheme benefits from the advancement in the optimization of neural-network modeling, which allows us to carry the supervised learning task conveniently under one caveat (a reasonable neural-network architecture, in our case LSTM, and various tuning parameters). In addition, it is found from our numerical tests that the model performance is insensitive to different choices of neural-network hyperparameters such as the input chain length and hidden state size, implying robust prediction skills of the model framework.

While this result is encouraging, we only view this work as a first step. Particularly, this work only focuses on the L-96 system with homogeneous statistical dynamics. This assumption simplifies the closure model as the covariance matrix naturally becomes diagonal, so we only need to close the diagonal variances and their reduction. A more important and challenging direction is to extend the proposed ML approach to non-homogeneous statistical dynamics, which involve nontrivial off-diagonal covariance components. Besides the curse of dimension problem (the dynamical equation of the covariance matrix has $N^2$-terms), a direct closure on the covariance statistics may not preserve positive definite-ness under the machine learning prediction. To overcome these two issues, one possibly needs to consider closing the dynamical equation for the fluctuation components $Z_i$ in \eqref{eq:spec_expansion} directly, extending the idea from \cite{sapsis2013blending} with ML model, which is part of our future work.

\section*{Acknowledgment}

The research of J.H. was partially supported under the NSF grants DMS-1854299, DMS-2207328, and the ONR grant N00014-22-1-2193. This research of D.Q. is partially supported by the Office of Naval Research N00014-19-1-2286.


\bibliographystyle{plain}  

\end{document}